\documentclass[journal,twoside,web]{ieeecolor}
\usepackage{tmi}
\usepackage{cite}
\usepackage{amsmath,amssymb,amsfonts}
\usepackage{algorithmic}
\usepackage{graphicx}
\usepackage{textcomp}
\usepackage{booktabs}
\usepackage{comment}
\usepackage{multirow}
\usepackage{makecell}
\usepackage{mathrsfs} 
\usepackage{adjustbox}
\usepackage{diagbox} 
\usepackage{hyperref}

\def\BibTeX{{\rm B\kern-.05em{\sc i\kern-.025em b}\kern-.08em
    T\kern-.1667em\lower.7ex\hbox{E}\kern-.125emX}}
\hypersetup{
	colorlinks=true,   
	linkcolor=blue,    
	citecolor=green,   
	urlcolor=red,      
	pdftitle={Your Document Title},  
	pdfauthor={Your Name}
}

\begin{document}
\title{Energy-Threshold Bias Calculator: A Physics-Model Based Adaptive Correction Scheme for Photon-Counting CT}
\author{Yuting~Chen, Yuxiang~Xing, Li~Zhang, Zhi~Deng and Hewei~Gao*
\thanks{This project was supported in part by the National Key R\&D Program of China under grant No.2022YFE0131100 and No.2024YFC2417500, and in part by Beijing Natural Science Foundation (No. L252021). \textit{(Corresponding author: Hewei Gao)} }
\thanks{The authors are with the Department of Engineering Physics, Tsinghua University, Beijing 100084, China, and are also with the Key Laboratory of Particle and Radiation Imaging, Ministry of Education, Tsinghua University, Beijing 100084, China (e-mail: 
	hwgao@tsinghua.edu.cn). }
}

\maketitle

\begin{abstract}

Photon-counting detector based computed tomography (PCCT) has greatly advanced in recent years. However, spectral inconsistency, referring to inter-pixel variations in detected counts per energy bin, can easily leads to ring or band artifacts and inaccuracies in CT reconstructed images. 
This work proposes a novel physics-model based method to correct for spectral inconsistency by modeling it through two terms: (1) a fixed spectral skew term (energy threshold-independent filtration function) determined at a given energy threshold, and (2) a variable energy-threshold bias term that can be directly calculated by using our spectral model as the threshold changes.
After the two terms being computed out in the calibration stage, they will be incorporated into our spectral model to adaptively generate the spectral correction vectors as well as the material decomposition vectors if needed, pixel-by-pixel for PCCT projection data.
Using a minimum set of parameters with explicit physics meaning, such an energy-threshold bias calculator (ETB-Cal) has advantages of computational efficiency, robustness in implementation, and convenience with no need of X-ray fluorescence materials in calibration.
To validate our method, both numerical simulations and physical experiments using multiple phantoms were carried out on a tabletop PCCT system, with preliminary results showing a significant reduction in non‑uniformity, from 29.3 to 5.8 HU for Gammex multi-energy phantom versus no correction (comparatively, 8.3 HU was achieved by a polynomial-involving model‑based approach with no explicit modeling and calculating of energy threshold bias but more calibration data required), and from 27.9 to 3.2 HU for the Kyoto head phantom.

\end{abstract}


\section{Introduction}
\label{sec:introduction}

\IEEEPARstart{S}{pectral} computed tomography (CT) imaging has advantages of eliminating beam hardening artifacts, providing information on material composition, reducing metal artifacts, making CT more quantitative\cite{Flohr2020review,danielsson2021review}. As a breakthrough technology, photon-counting detector (PCD)-based spectral CT has attracted significant research attention and prompted extensive investigations in recent years\cite{Willemink2018review,yu2016photoncounting}. Different from the energy-integrating detectors used in conventional spectral CT, PCDs directly counting photons with multiple energy bins simultaneously\cite{ballabriga2021review,Shikhaliev2011review}. 
With smaller pixel size and optimized energy thresholding that eliminates electronic noise, PCCT systems achieve higher spatial resolution and improved signal-to-noise ratio at reduced radiation doses, while enabling better material differentiation and quantitative K-edge imaging\cite{RN45,Mohamed2021kedge,Roessl2011kedge}.

However, there still exists serious physical and technical challenges to overcome in PCCT, such as the pulse pileup, charge sharing, bias in energy threshold, and other non-ideal effects\cite{Rajendran2021review,Rajbhandary2018calibration,Gimenez2011chargesharing,Iniewski2007chargesharing,Kafaee2020pileup}. These effects not only cause quantum losses or readout bias, but also change the characteristics of the signal pulses in each individual detector pixel, making PCCT vulnerable to artifacts and unstable in many cases\cite{ballabriga2021review}. 
This study focuses on energy threshold bias, a critical factor contributing to spectral inconsistency in PCCT.

In PCCT, detectors identify and count individual X-ray photons that exceed user-defined energy thresholds. An electronic discriminator compares each photon-induced pulse to digital-to-analog converter (DAC)-defined reference voltages, establishing energy bins based on the positive correlation between pulse amplitude and photon energy. However, threshold bias originates from pixel-to-pixel electronic variations in both detector fabrication and pulse processing circuitry\cite{danielsson2021review,Flohr2020photoncounting,ballabriga2021review}.
Without proper correction, threshold bias degrades energy resolution and introduces severe spectral inconsistencies, resulting in artifacts in reconstructed CT images and inaccuracies in material discrimination\cite{Taguchi2017review,Behbahani2020pileup}. 
Liu et al. reported that, in a CdTe PCD, the measured non-uniformity (pixel-wise standard deviation of flat-field counts) was 358 and 184 counts per pixel at threshold deviations of 2.44 and 0.34 keV, respectively \cite{Liu_2025}. Similarly, Li et al. demonstrated up to a threefold degrading in SNR and CNR in reconstructed images of Gd-doped phantoms when compared with energy-calibrated results\cite{li2016feasible}.

For energy-threshold bias, numerous approaches have been investigated in the literature\cite{Panta2015calibration,Ding2014xrfR,Ge2017kedge,Cammin2014distortions,Vespucci2019RobustEcal,alvarez2011estimator,sidky2022spekcali,lee2018spekdisto}, which can be grouped mainly into three categories, including monoenergy-based ones such as using the X-ray fluorescence materials, gamma-ray isotopes, voltage-peak, k-edge absorption or synchrotron beams to directly calibrate the DAC voltage values to ensure its corresponding energy threshold matches the monoenergy\cite{Panta2015calibration,Ding2014xrfR,Ge2017kedge}. Phantom-based measurement ones where a mapping from biased PCCT data to their corresponding nominal values is directly established for correction\cite{alvarez2011estimator}, and model-based ones which usually consist of both calibration and correction by using a physical or mathematical model of the PCCT data\cite{sidky2022spekcali,lee2018spekdisto,Cammin2014distortions}. 
Each of them has its own strengths and limitations\cite{rodesch2023comparison}. While monoenergetic X-ray fluorescence (XRF) is widely adopted for its simplicity, it suffers from the difficulty in obtaining the required XRF materials and a very low photon output. which leads to long data collection time and a tedious process in practice. For a phantom-based measurement method, it is quite straightforward to implement but its correction performance is often restricted to the specific phantom used, limiting its applicability.

Model-based calibration approaches offer distinct advantages in terms of procedural simplicity and computational efficiency, enabling effective spectral correction for arbitrary objects\cite{fangwei2020NN}. However, their performance critically depends on the accurate modeling of PCCT imaging chains, which requires comprehensive physical understanding. 
Among existing methods, Sidky et al. developed a calibration model for the spectral response and polynomial flux dependence\cite{sidky2022spekcali}. Although parameter-efficient, this method relies on an empirical polynomial model that implicitly combines multiple physical effects without explicitly modeling threshold bias. Lee et al. proposes a three-step estimator using low-order polynomial approximation of x-ray transmittance to estimate basis line-integrals in PCCT with modeled spectral distortion \cite{lee2018kedge,lee2018spekdisto}. This method, though theoretically valid, relies heavily on pre-calibration and suffers performance loss out of the limits of the calibration ranges. 
Both methods used polynomial approximations for spectral distortion and/or flux calibration, which are straightforward and commonly used in practice but their performance usually rely on the selection of orders and terms of polynomials, and could cause oscillations in fitting due to the Runge phenomenon \cite{runge1901empirische}. Also, neither method explicitly models energy threshold bias, which is a key focus of our study.

In this work, we present a robust and adaptive physics-model based energy-threshold bias calculator (ETB-Cal) by modeling spectral inconsistencies through a spectral skew term and an energy-threshold bias term, which enables an analytical characterization and adaptive correction for ETBs across detector pixels, making the commonly used but practically tedious XRF based calibrations unnecessary. To validate our method, both numerical simulations and physical experiments were carried out on a tabletop PCCT system. In our previous work\cite{chen2024physics}, some results focusing exclusively on the spectral skew term were presented, while in this study, we refined the spectral model and provided a rigorous formulation of the mathematical solution. Additionally, more experiments have been conducted, including quantitative calculations of simulated key parameters, material decomposition of an anthropomorphic head phantom, and comparisons with two existing model-based methods, as well as with the XRF-based ETB measurements. The main contributions of this work are summarized as follows:
\begin{itemize}
	\item Physics-based threshold modeling: The model explicitly characterizes spectral inconsistency across energy thresholds bias and spectral skew term.
	\item Efficient correction method: A parameter-efficient design allows fast multi-threshold inconsistency computation without sacrificing accuracy.
	\item Calibration process simplification: Minimal calibration workflow with no need of XRF materials reduces operational complexity in practical applications.

\end{itemize}

The rest of this paper is organized as follows. Section II describes the details of our proposed ETB-Cal algorithm with two-term spectral modeling. Section III and Section IV present the experimental settings and the corresponding results, respectively. Section V concludes our study with a brief discussion.

\section{Method}

\subsection{Novel Spectral Modeling with Two-Term Factorization}
\begin{figure}[!t]
	\centering
	\centerline{\includegraphics[width=1\columnwidth]{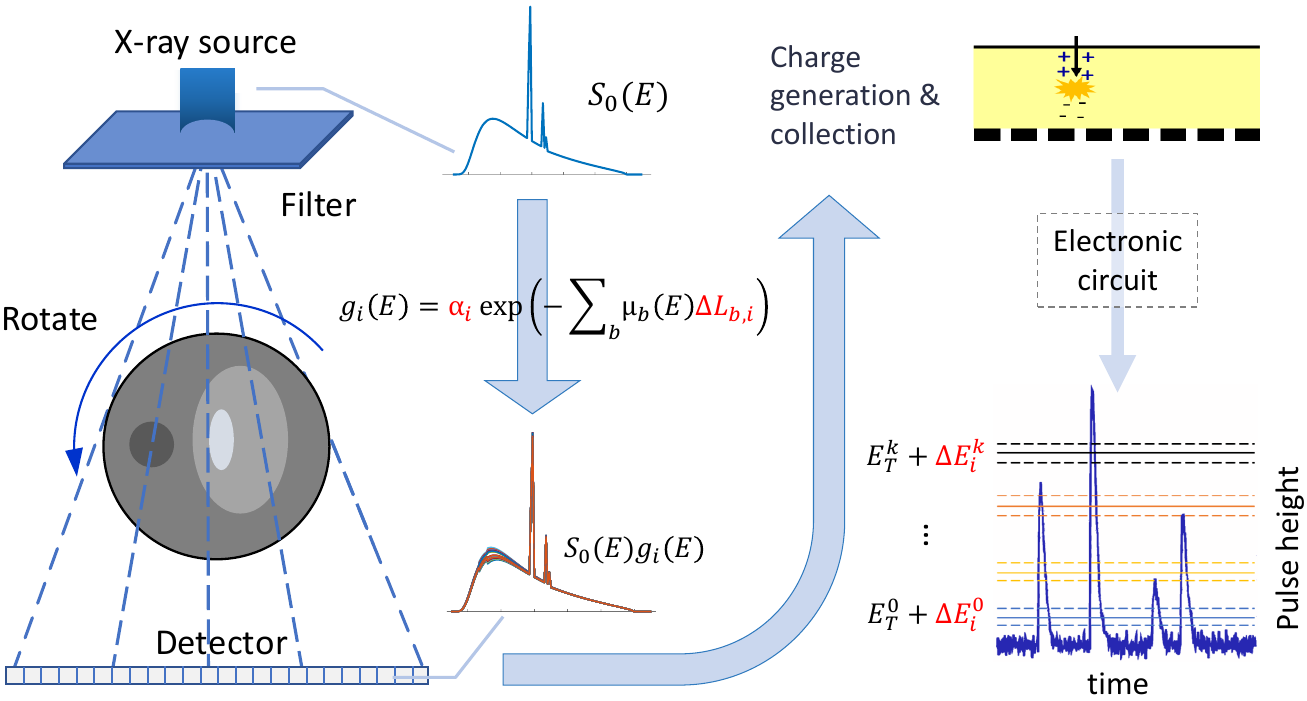}}
	\caption{{Schematic diagram of PCCT data acquisition mechanism, which illustrates the sources of spectral inconsistency in the proposed physics-based two-term ($\Delta E^{k}_{i}$ and $g_{i}(E)$) factorized spectral model and the specific forms of the model parameters. $E^{k}_{T}$ is the $k$-th nominal energy threshold with bias $\Delta E^{k}_{i}$ varying across different detector pixels, $g_{i}(E)$ is the spectral skew term determined by thickness deviations ($\Delta \textit{L}_\textit{b,i}$) of different material combinations.}} 
	\label{fig:PCCT-Diagram}
\end{figure}

In PCCT, PCD records the photons whose deposited energies fall into specified energy ranges. The photon count $\lambda^{k}_{i}$ at the $i$-th pixel above the $k$-th energy threshold \(E_{{T}}^k\) can be modeled as,

\begin{equation}
\lambda^{k}_{i}=\int_{E_{{T}}^k}^\infty{\int_{0}^\infty{{S_{0i}(E)H_i(E)R_i(E^\prime;E)}{\rm d}E}{\rm d}E^\prime}, 
\label{eq:pcdModel-Standard} 
\end{equation}
with
\begin{equation}
    H_i(E) = {\rm e}^{-\sum \mu_n(E)L_{n_{i}}}.
    \label{eq:Obj-Atten} 
\end{equation}

\noindent where, $S_{0i}(E)$ is the incident X-ray spectrum, available via Monte Carlo simulation or measurement; $H_i(E)$ describes object attenuation, determined by material coefficients $\mu_n(E)$ and path lengths $L_{n_i}$; 
${R}_i(E^\prime; E)$ is the energy response of the $\textit i$-th detector pixel.
Such an idealized formulation in Eq. \eqref{eq:pcdModel-Standard} does not explicitly incorporate some key factors that cause spectral inconsistencies in detector outputs in real applications.

As illustrated in Fig.~\ref{fig:PCCT-Diagram}, in practice, two terms related to spectral inconsistency can be commonly encountered here. One is associated with the variations resulting from different spectral skew among different pixels, and the other is related to the variations in energy thresholds. Accordingly, a novel, unified, and two-term factorized PCCT spectral model can be established as

\begin{equation}
\lambda^{k}_{i}=\int_{E^{k}_{T}+\Delta E^{k}_{i}}^\infty{\int_{0}^\infty{S_{0i}(E)g_{i}(E)H_i(E){R}_i(E^\prime;E){\rm d}E}{\rm d}E^\prime},	
\\[5pt]
\label{eq:TTF-spectralModel} 
\end{equation}
\begin{equation}
	\begin{split}
		g_{i}(E)=\alpha_i \cdot \rm{\exp}(- \sum_{\textit{b}}\mu_\textit{b}(\textit{E})\Delta \textit{L}_\textit{b,i} )
	\end{split}
\label{eq:giE expression} 
\end{equation}
in which $i$ and $k$ represent detector pixel index and energy threshold index, the spectral skew term is modeled by a filtration function $g_i(E)$ and energy threshold variations are captured through a bias term $\Delta E_i^k$ that adjusts the nominal threshold $E_T^k$. Physical meanings of the two terms are as follows.

$\Delta E^{k}_{{i}}:$ During data acquisition, in general, a uniform nominal threshold value will be applied across all pixels. 
While the actual threshold deviates from the nominal value and varies across pixels due to the electronic imperfections such as thermal drift and noise. We used $\Delta E^{k}_{i}$ to denote this energy threshold bias (ETB).

$g_{i}(E):$ In practical applications, deviations are commonly observed in $S_{0i}$, necessitating compensation via $g_i(E)$. While the spectral filtration function is a common tool to modify the hardness of spectrum, according to material decomposition theory \cite{alvarez1976energy}, spectral filtration can be achieved by combining different basis materials, with the attenuation behavior represented using two kinds of non-K-edge materials and the K-edge material itself (if involved). Thus, $g_i(E)$ can cover majority of practical filtration scenarios by adjusting material categories and thickness deviations ($\Delta L_{b,i}$). The coefficient $\alpha_i$ ($\approx 1$) represents minor flux changes across pixels.

\begin{figure}[!t]
	\centering
	\centerline{\includegraphics[width=1\columnwidth]{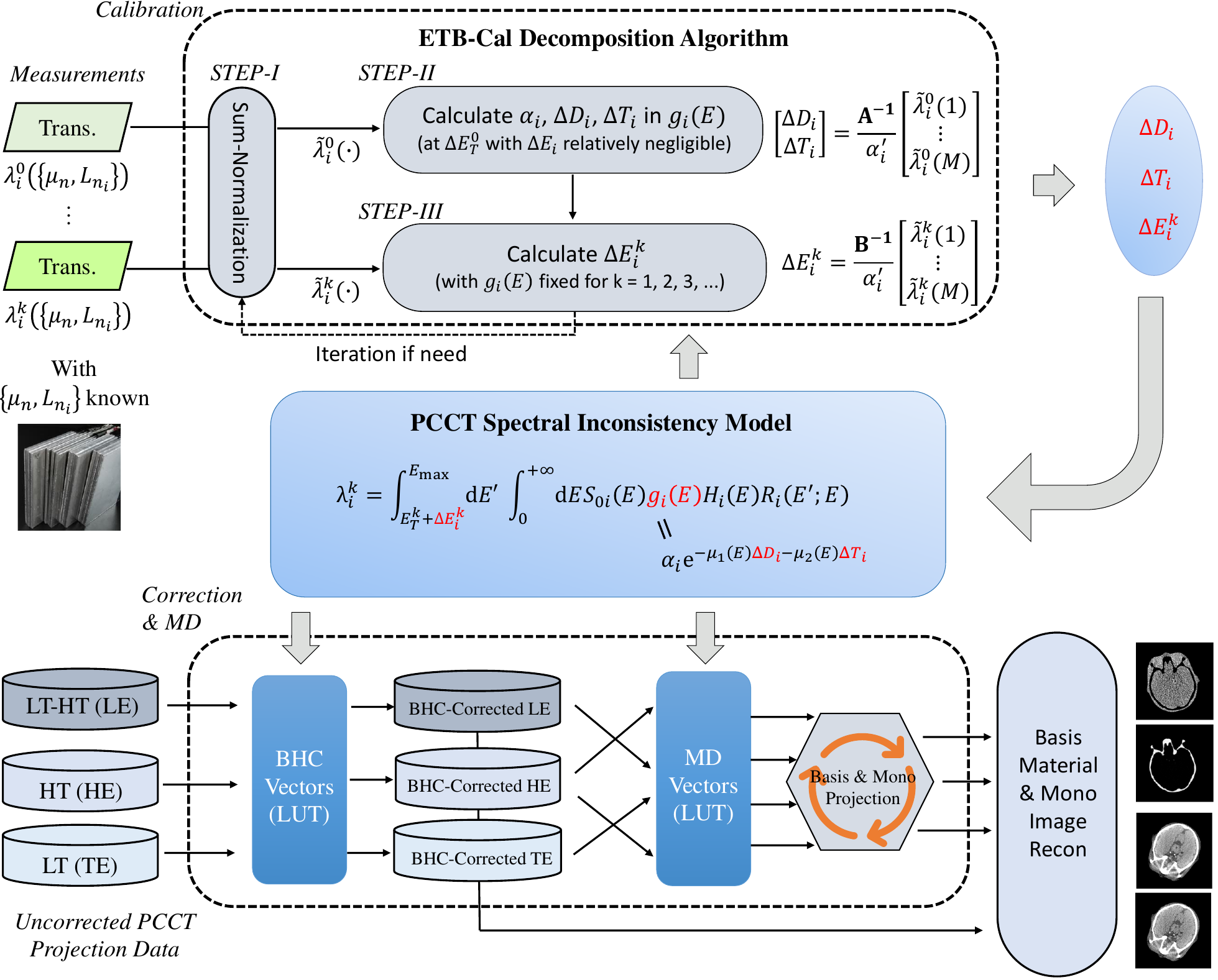}}
	\caption{{The overall framework of the spectral inconsistency correction. Based on the proposed spectral model, $\alpha_{i}$, $\Delta D_{i}$, $\Delta T_{i}$, $\Delta E^{k}_{i}$ can be determined through a series of attenuation measurements with known $\mu _{n}$ and $L _{n}$. By applying these parameters to beam hardening correction process and material decomposition process, we can correct the ring and band artifacts resulting from the spectral inconsistency in the reconstruction images.}}
	\label{fig:framework}
\end{figure}

\subsection{Key Linearization Assumptions for Model Simplification}\label{sec:Key Linearization Assumptions}
To simplify the model, we highlight two key assumptions. 

Assumption 1: The energy threshold bias $\Delta E^{k}_{i}$ in Eq.~(\ref{eq:TTF-spectralModel}) is sufficiently small, making the integral over $[E^{k}_{T}, E^{k}_{T}+\Delta E^{k}_{i}]$ can be linearly approximated by evaluating the integrand at $E^{k}_{T}$ and scaling with $\Delta E^{k}_{i}$, as formalized in 
\begin{equation}
	\begin{split}
		&\Delta E^{k}_{i}\approx \frac{\int_{E^{k}_{T}}^{E^{k}_{T}+\Delta E^{k}_{i}}{S_{0i}(E)g_i(E)H_i(E){R}_i(E^\prime;E)}{\rm d}E^\prime}{S_{0i}(E)g_i(E)H_i(E){R}_i(E_T^k;E)}.
	\end{split}
\label{eq:ETB linear approx} 
\end{equation}

Assumption 2: In our implementation, the spectral skew term \( g_i(E) \) can be approximated using the detector material and the pre-filter material as follows:
\begin{equation}
	\begin{split}
		g_{i}(E)&\approx\alpha_i\cdot{\rm e}^{- \mu_1(E)\Delta D_{i} - \mu_2(E) \Delta T_{i}}\\
		&\approx \alpha_i(1 - \mu _1(E) \Delta D_i - \mu _2(E) \Delta T_i).
	\end{split}
	\label{eq:giE linear approx} 
\end{equation}
Here, $\mu_1(E)$, $\mu_2(E)$ and $\Delta D_i$, $\Delta T_i$ are the linear attenuation coefficient and thickness variations of cadmium telluride (CdTe) and aluminum (Al). The approximation is a compromise between increasing categories of basis materials in Eq. \eqref{eq:giE expression} (which amplifies decomposition errors) and maintaining sufficient spectral characterization fidelity, validated in section \ref{Results}. Moreover, it assumes that the $ \mu_1(E)\Delta D_i + \mu_2(E)\Delta T_i $ across all pixels has an overall insignificant effect over the entire energy of interest, thereby keeping the error from the first-order Taylor expansion within an acceptable range.

Taking advantage of these two approximations, we can linearize and simplify our spectral model in Eq.~\eqref{eq:TTF-spectralModel} as
\begin{equation}
	\begin{aligned}
		\lambda^{k}_{i}&\approx \alpha_i\left [ \begin{matrix}
			1 
			&-\Delta D_i 
			&-\Delta T_i
		\end{matrix} \right ]
	 \left [ \begin{matrix}
		A_{1i}^k &-B_{1i}^k \\
		A_{2i}^k &-B_{2i}^k \\
		A_{3i}^k &-B_{3i}^k 
	\end{matrix} \right ]
 \left [ \begin{matrix}
	1 \\
	\Delta E_i^k 
\end{matrix} \right ]
	\end{aligned}
\label{eq:lambda rewrite} 
\end{equation}
\noindent where,

\begin{equation} 
	\begin{aligned}
		&A_{li}^k=\int_{E^{k}_{T}}^\infty{{F_{li}(E^\prime)}{\rm d}E^\prime},~B_{li}^k=F_{li}(E_i^k),~(l=1,2,3)\\
		&F_{1i}(E^\prime)=\int_{0}^\infty{S_{0i}(E)H_i(E){R}_i(E^\prime;E){\rm d}E}\\
		&F_{2i}(E^\prime)=\int_{0}^\infty{S_{0i}(E)\mu _1(E)H_i(E){R}_i(E^\prime;E){\rm d}E}\\
		&F_{3i}(E^\prime)=\int_{0}^\infty{S_{0i}(E)\mu _2(E)H_i(E){R}_i(E^\prime;E){\rm d}E},	
	\end{aligned}
\label{eq:item abbrs}
\end{equation}

With the analytical expressions for \( S_{0i}(E)\), \( H_i(E) \), and \( R_i(E^\prime;E) \) provided, the value of \( F_{li} \) becomes fully determined. The only unknown parameters in the model are $\alpha_{i}$, $\Delta D_{i}$, $\Delta T_{i}$, and $\Delta E^{k}_{i}$. 

To solve for these unknown parameters, it is necessary to generate different detector-recorded spectra which can be achieved by altering \( S_{0i}(E) \) or modifying \( H_i(E) \). Common approaches include varying kVp or using filters with known \{$\mu_n(E)$, $L_{n_i}$\} to enable the transmission measurement of the  \( S_{0i}(E) \). The latter approach is adopted in this work. 

The two terms introduced in Eq. \eqref{eq:TTF-spectralModel} are complementary in the model, and there is no one-step analytical solution clearly. Iterative solving for the parameters in two terms is doable, but it can easily suffer in robustness and convergence issues, and may require excessive computational demands in practice as in the approximate direct inversion of Eq. \eqref{eq:lambda rewrite} (refer to Appendix \ref{Direct Inversion Method} for details). 
To overcome these limitations, we proposed ETB-Cal, a physics-based decomposition algorithm offering both robustness and computational efficiency as follows.

\subsection{ETB-Cal Decomposition Algorithm}\label{sec:ETB-Cal Decomposition Algorithm}
As shown in Fig. \ref{fig:framework}, the ETB-Cal decomposition algorithm comprises sum-normalization on $M$ sets of $\lambda _i^k$ measurements, calculation of $g_i(E)$, and calculation of $\Delta E_i^k$.

\subsubsection{Pre-processing for the Raw Data}
This refers to $\textit{STEP-I}$ in Fig. \ref{fig:framework}. Let $m = 1, 2, \dots, M$ represent the number of measurements with different spectra. 
For the $m$-th set of counts (in our work acquired after X-ray transmission through flat filters), we define the normalized photon counts as 

\begin{equation}
	\begin{aligned}
		\tilde{\lambda}^{k}_{i}(m) = \frac{\lambda^{k}_{i}(m)}{\sum_{j=1}^{N} \lambda^{k}_{j}(m)}	
	\end{aligned}
\label{eq:self norm} 
\end{equation}
where $i$ and $j$ denote the detector pixel index, $N$ denotes the total number of pixel in selected detector area.

This sum-normalization operation in Eq. \eqref{eq:self norm} eliminates the need for absolute photon counts ($\lambda_i^k$) and focuses on relative errors between measurements and model predictions, with the constraint $\sum_{i=1}^{N} \tilde{\lambda}^{k}_{i}(m) = 1$ improves the stability of the solution process.

\begin{figure}[!t]
	\centerline{\includegraphics[width=0.9\columnwidth]{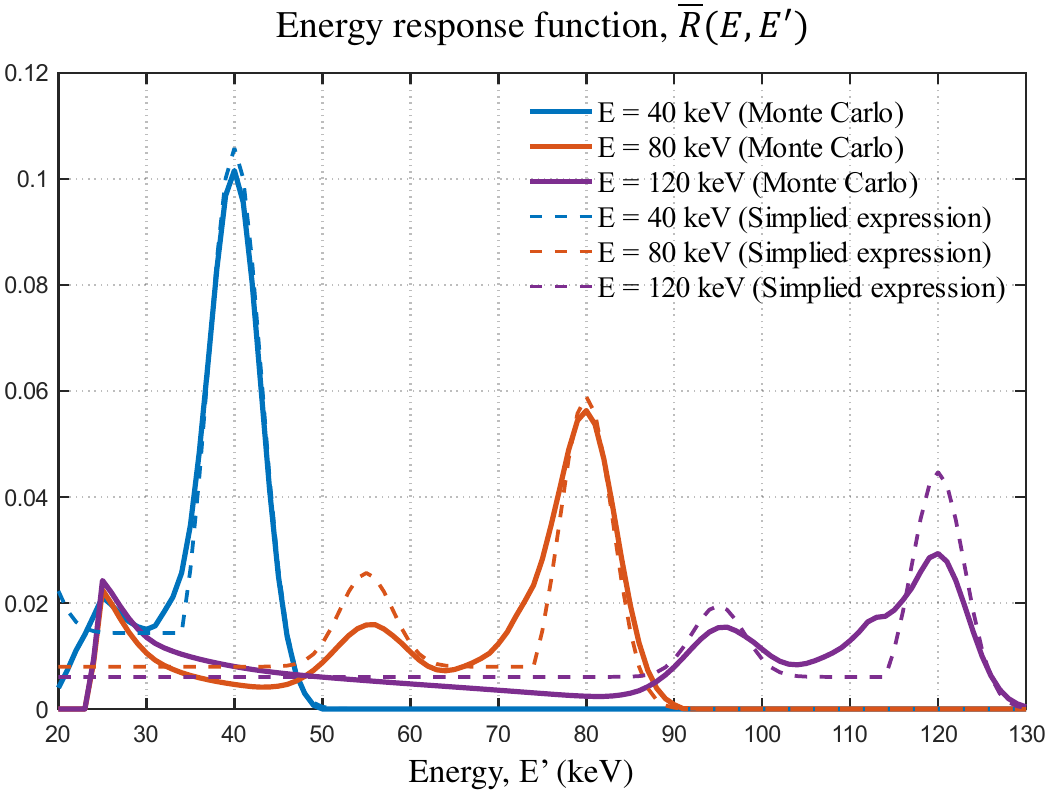}}
	\caption{{Energy response functions $\bar{R}(E,E^\prime)$, when E = 40, 80, and 120 keV. Dashed line: schematic of the energy response function, which incorporates three principal components: (1) the primary photopeak resulting from incident photon energy deposition, (2) the characteristic fluorescence escape peak, and (3) a modeled constant Compton continuum. Solid line: The energy response function, generated by our Monte Carlo simulations, used in our experiments. Based on the decoupling of the energy threshold bias (ETB) term and spectral skew term $[\Delta E^{k}_{i}$ and $g_{i}(E)]$, a reasonably appropriate base energy threshold $ E^{0}_{T}$ can be selected in the low-energy range.}}
	\label{fig:energy response}
\end{figure}

\subsubsection{Feasibility Analysis of Decoupling the Two Terms}\label{sec:Feasibility Analysis of Decoupling}

Fortunately, our model allows the two terms to be decoupled, enabling $g_{i}(E)$ to be separated from $\Delta E^{k}_{i}$ based on the following considerations. 

On the one hand, the relation between the output voltage threshold $V_{T}$ and the
detected energy threshold $E_{T}$ is linear and characterized by a gain $k$ and offset $b$ 
\begin{equation}\label{eq:E-V relation} 
	E_{T}=kV_{T}+b.
\end{equation}

Regarding the fact that the global $E_T$-$V_T$ calibration (based on Eq.~\eqref{eq:E-V relation}) can yield pixel-specific $(k_i,b_i)$ resulting from non-ideal factors, 
the variance of $\{\Delta E_i\}$ can be written as
\begin{equation}\label{eq:dE} 
	\begin{split}
	\mathbb{V}\mathrm{ar}\{\Delta E_i\}=(V_T)^2\mathbb{V}\mathrm{ar}\{k_i\} +\mathbb{V}\mathrm{ar}\{ b_i\},
	\end{split}
\end{equation}

\noindent in which, $\mathbb{V}\mathrm{ar}\{\Delta E_i\}$ exhibits a monotonic increase with threshold voltage $V_T$.

Therefore, it is more reasonable to assume that \( \Delta E^{k}_{i} \) is fairly small at lower \( E_T \) (\( E_T^0 \)) where the distribution range of $\Delta E^{k}_{i}$ for all pixels is relatively narrower, then calculating \( g_{i}(E) \) alone is feasible.
The spectral inconsistency due to the energy-threshold bias at $ E^{0}_{T}$ can be considered to be overwhelmingly encompassed within $g_{i}(E)$ (details seen in appendix \ref{giE compensation}).

On the other hand, as shown in Fig. \ref{fig:energy response} and \cite{danielsson2021review,Cammin2014distortions}, the PCD energy response exhibits a slowly varying region (Compton plateau) below the incident photon energy (for details on energy response simulations, refer to Section \ref{sec:experiment Setup}). In this Compton-dominated regime, small energy threshold biases ($\Delta E^{k}_{i}$) cause only minor adjustments to the integrated response curve area. Consequently, incorporating $\Delta E^{k}_{i}$ into $g_{i}(E)$ results in errors that are practically insignificant in PCCT system.

Based on these observations, we proposed a two-step decoupling strategy to solve for the unknown terms in Eq. \eqref{eq:TTF-spectralModel}, comprising (1) derivation of $g_{i}(E)$ at the reference threshold $E_{T}^{0}$ and (2) computation of $\Delta E^{k}_{i}$ for the remaining energy thresholds.

\subsubsection{Calculation for Spectral Skew Term}\label{sec:Calculation for Spectral Skew Term}

This refers to $\textit{STEP-II}$ in Fig. \ref{fig:framework}. Based on the per-pixel assumption $\mu_1 \Delta D_i + \mu_2 \Delta T_i \approx 0$ used in Eq. \eqref{eq:giE linear approx}, we derived
\begin{equation}
	\begin{aligned}
		\tilde{\lambda}^{0,model}_{i}(m)\approx \frac{\alpha_i[A_{1i}^0(m)-\Delta D_iA_{2i}^0(m)-\Delta T_iA_{3i}^0(m)]}{\sum_{i=1}^{N} \alpha_i A_{1i}^0 (m)}.
	\end{aligned}
	\label{eq:normed lambda} 
\end{equation}
from Eq. \eqref{eq:self norm}, thereby removing $\sum_{i=1}^{N} \alpha_i [\Delta D_iA_{2i}^0(m)+\Delta T_iA_{3i}^0(m)]$ from the denominator.
 
Under the assumptions: (1) In Eq. \eqref{eq:item abbrs}, $S_{0i}(E)$, $R_i(E';E)$ are identical across pixels (spectral inconsistency is separately captured by $g_i(E)$ and $\Delta E_i^k$) and $H_i(E)$ is nearly constant (source to detector distance is much larger than the detector width, making the attenuation length difference in flat filters measurements for calibration can be ignored). (2) $\Delta E_i^{0}$ is small. With (1), $A_{li}^{0}$ varies little across pixels. (3) $\alpha_i$ reflects small flux changes across pixels, so $\alpha_i \approx 1$.

We further assumed the existence of an equivalent coefficient $\overline{\alpha}$ satisfying the following relation for a simpler formulation:
\begin{equation}
	\begin{aligned}
		\overline{\alpha}\sum_{i=1}^{N} A_{1i}^0 (m)\approx \sum_{i=1}^{N} \alpha_i A_{1i}^0(m)
	\end{aligned}
\label{eq:alpha extract} 
\end{equation}

\noindent Let $\alpha^\prime_i=\frac{\alpha_i}{\overline{\alpha}}$, and
\begin{equation}
	\tilde v_i^0(m)=
	\begin{aligned}
		\frac{1}{{\sum_{i=1}^{N} A_{1i}^0 (m)}} \left [ 
		\begin{matrix}
			{A_{1i}^0(m)}&-{A_{2i}^0(m)} &-{A_{3i}^0(m)}\\
		\end{matrix}
		\right ],
	\end{aligned}
\label{eq:A vecter} 
\end{equation}
by repeating $M$ sets of X-ray transmission measurements, one can derive the following linear equations from \{$\tilde{\lambda}^{0,model}_{i}(m) = \tilde{\lambda}^{0,measurement}_{i}(m)$\}$_{m=1,2...,M}$,

\begin{equation} 
	\left [ \begin{matrix}
		\tilde \lambda^{k,meas.}_i(1) \\
		\tilde \lambda^{k,meas.}_i(2) \\
		\vdots\\
		\tilde \lambda^{k,meas.}_i(M) \\
	\end{matrix} \right ]
	\approx \alpha^\prime_i \left [ 
	\begin{matrix}
		\tilde v_i^k(1)\\
		\tilde v_i^k(2)\\
		\vdots \\
		\tilde v_i^k(M) \\
	\end{matrix}
	\right ] \left [ \begin{matrix}
		1 \\
		\Delta D_i \\
		\Delta T_i \\
	\end{matrix} \right ]
\label{eq:matrix equation}
\end{equation}

Once $\alpha^\prime_i$, $\Delta T_i$, and $\Delta D_i$ in Eq. \eqref{eq:matrix equation} are obtained, only $\Delta E^{k}_{i}$ requires threshold-dependent updates during the spectral inconsistency correction. In our work, Eq. \eqref{eq:matrix equation} is solved by Levenberg–Marquardt least-squares minimization \cite{nocedal2006numerical}.

\subsubsection{Energy-threshold Bias Calculator}

This refers to $\textit{STEP-III}$ in Fig. \ref{fig:framework}. With $\alpha^\prime_i$, $\Delta T_i$, and $\Delta D_i$ determined, $\Delta E^{k}_{i}$ can be analytically calculated by combining Eqs.~\eqref{eq:lambda rewrite} and \eqref{eq:self norm} as
\begin{equation}
	\begin{aligned}
		&\Delta E_{i}^{k}=\frac {-(\frac{\tilde{\lambda_i^k}}{{\alpha_i}\prime 
			}\sum_{i=1}^{N} A_{1i}^k)+ (A_{1i}^k - \Delta D_i A_{2i}^k - \Delta T_i A_{3i}^k) }{B_{1i}^k - \Delta D_i B_{2i}^k - \Delta T_i B_{3i}^k}	  
	\end{aligned} 
\label{eq:ETB-Cal} 
\end{equation}

Eq.~(\ref{eq:ETB-Cal}) intuitively reflects the characteristic of this method as a calculator, enabling direct computation of $\Delta E_{i}^{k}$ through a closed-form expression with all variables known. When incorporating $M$ sets of transmission measurements, $\Delta E_{i}^{k}$ is optimally determined through least-squares fitting as it becomes an over determined problem.

In practice, to satisfy the linearization assumption for $\Delta E_{i}^{k}$ (Section~\ref{sec:Key Linearization Assumptions}), its magnitude must be kept small. The true value of $\Delta E_{i}^{k}$ can be approximated by repeated calculations. Within this framework, we update the threshold as $E_T^{k}(\mathrm{repeat}_{n+1}) = E_T^{k}(\mathrm{repeat}_{n}) + \Delta E_{i}^{k}(\mathrm{repeat}_{n+1})$.

\subsection{Benefits of the Spectral Model and ETB-Cal Algorithm}

Using the two-term factorization model, the function $g_{i}(E)$, derived at a specific low threshold $E_{T}^0$, can be universally applied to all higher thresholds. As a result, for the $k$-th threshold (where $k=1,2,3,…$), only one unknown parameter $\Delta E^{k}_{i}$ needs to be computed. Then spectral inconsistencies can be adaptively corrected through beam hardening correction and material decomposition vectors as shown in Fig.~\ref{fig:framework}.

This spectral model and ETB-Cal algorithm significantly enhances the applicability of spectral inconsistency correction while reducing computational demands, providing the following advantages.
1) Once the model parameters are determined, they can be applied universally to any object scanned;
2) The model is parameter-efficient, enabling a rapid calculation of inconsistency metrics across multiple thresholds;
3) Only a minimal amount of data is required for adaptive online calibration;
4) It bypasses the need for the complex XRF calibration process, simplifying its implementation in practice.

\subsection{Spectral Inconsistency Correction}\label{sec:Spectral Inconsistency Correction}
Following calibration with measurements acquired under different effective spectra, the resulting values of $\Delta D_{i}$, $\Delta T_{i}$, and $\Delta E^{k}_{i}$ are used to generate the beam hardening correction (BHC) vectors via

STEP-I: Select a range of material thicknesses (L’s); 

STEP-II: Set an effective energy ($\bar{E}$); 

STEP-III: Compute the monochromatic projection ($P_{mon}$) and the polychromatic projection ($P_{pol}$) using 
\begin{equation}
	P_{mon}(L) = \mu_o(\bar{E})L,
	\label{eq:BHC-solve1}
\end{equation}
\begin{equation}
	\begin{aligned}
		P_{pol,i}^k=-\ln \left( \frac{\int_{0}^\infty{{S_{\mathrm{eff},i}^k(E)g_i(E){\rm e}^{-\mu_o(E)L}{\rm d}E}}}{\int_{0}^\infty{{S_{\mathrm{eff},i}^k(E) g_i(E){\rm d}E}}} \right),
	\end{aligned}
	\label{eq:BHC-solve2}
\end{equation}
in which, $S_{\mathrm{eff}}(E)$ is the effective spectrum based on the detector’s energy response,
\begin{equation}
	\begin{aligned}
		S_{\mathrm{eff},i}^k(E) = \int_{E_{T}^k+\Delta E_i^k}^\infty{{S_{0i}(E){R}_i(E^\prime;E){\rm d}E^\prime}}.
	\end{aligned}
	\label{eq:Seff2}
\end{equation}

STEP-IV: Establish a lookup table that maps the polychromatic projection $\{P_{pol,i}^k\left(L\right)\}_{L=0}^{max}$ to its corresponding monochromatic one ($\{P_{mon}\left(L\right)\}_{L=0}^{max}$), with a polynomial fitting method\cite{hsieh2022computed}. Here, we express the obtained $j$-th order polynomial coefficients using the symbol \( \{d_j\} \).

Thus, projection $P_i^k$ obtained in practical measurements can be corrected through

\begin{equation}
	P_{\mathrm{corr},i}^k = \sum_{j} d_{j} (P_i^k)^j.
	\label{eq:BHC}
\end{equation}

\noindent  where, $i$ and $k$ represent the index of pixel and energy threshold, $P_{\mathrm{corr},i}^k$ is the projection with inconsistency corrected.

\subsection{Material Decomposition}
Projection-based material decomposition (MD) inherently relies on spectral information, necessitating pixel-specific decomposition mappings when spectral inconsistencies exist in the measurements.

In dual‑basis MD, ignoring K‑edge effects, two sets of projection data with distinct spectral information are required. Once determined, the parameters $\alpha_i^\prime$, $\Delta T_i$, $\Delta D_i$, and $\Delta E_i^k$ are substituted into Eq.~\eqref{eq:TTF-spectralModel} to generate MD attenuation vectors in a 2D format, following the same procedure used to create the BHC vectors but with two materials ($m_1, m_2$) and corresponding attenuation thicknesses $\rm{L_1^\prime s}, \rm{L_2^\prime s}$ as
\begin{equation}
	\begin{aligned}
		&P_{podm,i}^{k}(E,L_1,L_2)\\
		&=-\ln \left( \frac{\int_{0}^\infty{{S_{\mathrm{eff},i}^k(E)g_i(E){\rm e}^{-\mu_{m_1}(E)L_1-\mu_{m_2}(E)L_2}{\rm d}E}}}{\int_{0}^\infty{{S_{\mathrm{eff},i}^k(E) g_i(E){\rm d}E}}} \right).\\
	\end{aligned}
	\label{eq:MD solve}
\end{equation}
\noindent where the expression for ${S_{\mathrm{eff},i}^k(E)}$ is given in Eq. \eqref{eq:Seff2}.
	
For two selected $E_T^k$ ($k=k_1,k_2, E_T^{k_1}<E_T^{k_2}$), 
the MD coefficients $a_{pq,i}$ and $b_{pq,i}$ for $i$-th pixel are obtained by mapping $\{P_{podm, i}^{k_1}(E,L_1,L_2)\}$ and $\{P_{podm, i}^{k_2}(E,L_1,L_2)\}$ to the two monochromatic projections $\{\mu_{m_1}(\bar{E})L_1\}$ and $\{\mu_{m_2}(\bar{E})L_2\}$ using a polynomial function \cite{alvarez1976energy}.

Once $\{a_{pq,i}\}$ and $\{b_{pq,i}\}$ obtained, the basis material projections with inconsistency corrected can be generated through
\begin{equation}
	\begin{aligned}
	M_{1,i} = \sum_{pq} a_{pq, i} P_{{\rm {L}},i}^p P_{{\rm {H}},i}^q, \quad M_{2,i} = \sum_{pq} b_{pq, i}  P_{{\rm {L}},i}^p P_{{\rm {H}},i}^q,
	\end{aligned}
\label{eq:md}
\end{equation}
where, $P_{{\rm {L}},i}$ and $P_{{\rm {H}},i}$ represent measurements of the projections in the PCCT scan at two selected $E_T^{k}$ mentioned above (here we use $P_{{\rm {L}},i}$ and $P_{{\rm {H}},i}$ in place of $P_{i}^{k_1,meas.}$ and $P_{i}^{k_2,meas.}$ to avoid notational conflicts), $M_{1,i}$ and $M_{2,i}$ are the line integrals of the densities of two basis materials.

In our study, projection-domain MD using water and iodine as the basis materials was performed on a multi-energy phantom. While for head phantom, extremely low signal caused by bone structures significantly compromise decomposition accuracy. With the help of multi-material spectral correction (MMSC) method\cite{US9025815(B2),wang2023dual}, a post-reconstruction technique for conventional CT that estimates and corrects bone-induced beam hardening, we derived beam-hardening-free projections as 
\begin{equation}
	\begin{aligned}
		&P_{\rm {MMSC}} = P_t + \delta p(P_t, P_b).
	\end{aligned} 
	\label{eq:MMSC-P} 
\end{equation}

\noindent Here, \( P_t \) and \( P_b \) respectively represent total and bone projection, \( \delta p(P_t, P_b) \) refers to the bone-induced beam hardening error, which is typically approximated using a polynomial function as
\begin{equation}
	\begin{aligned}
		&\delta p(P_t,P_b)= \sum\limits_{i,j}c_{ij} P^i_t P^j_b	.  
	\end{aligned} 
	\label{eq:MMSC-dP} 
\end{equation}

After two sets of low- and high- energy beam hardening free projection are obtained. One can produce basis images \{$B_1,B_2$\} through image-domain material decomposition as
\begin{equation}
	B_1,B_2=\mathcal{M}^{-1}_{I}(\mathcal{R}(P_{\rm {MMSC}}^L, P_{\rm {MMSC}}^H)),
	\label{eq:image-domain MD} 
\end{equation}
where ${\mathcal{M}}^{-1}_{I}$ and ${\mathcal{R}}$ are image-domain material decomposition and reconstruction operator, respectively.

Also, virtual monoenergetic images (VMIs) at the target energy can be generated by weighting the basis images \{$B_1,B_2$\} using the mass attenuation coefficients of respective basis materials.

\subsection{Some Details of Comparative Study} \label{Some Details of Comparative Study} 
To reveal the performance of our proposed ETB-Cal method over other existing ones, we implemented four approaches as follows. (1) the direct inversion method (described in Appendix \ref{Direct Inversion Method}), (2) ring removal approach using windowed filtering of polar-transformed masked reconstructions \cite{sijbers2004reduction}, (3) the SFR method \cite{sidky2022spekcali} and (4) the BCT method\cite{lee2018spekdisto} (Detailed implementations of latter two were provided in Section \ref{Implementation of Two Existing Model-based Methods for Comparison} and Appendix \ref{Key implementation details of the two comparative model-based methods}).

We divided the four methods into two groups: the first two methods for preliminary comparisons, and the latter two recent model-based methods for more comprehensive comparisons. In Section \ref{sec:experiment-evaluations}, the implementations of recent model-based methods in simulation and experiment were put together in \ref{Implementation of Two Existing Model-based Methods for Comparison}. 
In Section \ref{Results}, we firstly presented results from ETB-Cal, direct inversion, and ring removal approaches in \ref{results: Simulated Abdominal Phantom} and \ref{results: Physical Experiments}.
Then, performance assessments of ETB-Cal against two existing model-based methods were presented to provide more comprehensive comparisons with the state of the art in Section \ref{results: Comparison with Two Existing Model-based Methods}.

\section{Experiment Setup and Evaluations}\label{sec:experiment-evaluations}
\subsection{Experimental Setup}\label{sec:experiment Setup}

\begin{figure}[!t]
	\centering
	\centerline{\includegraphics[width=0.85\columnwidth]{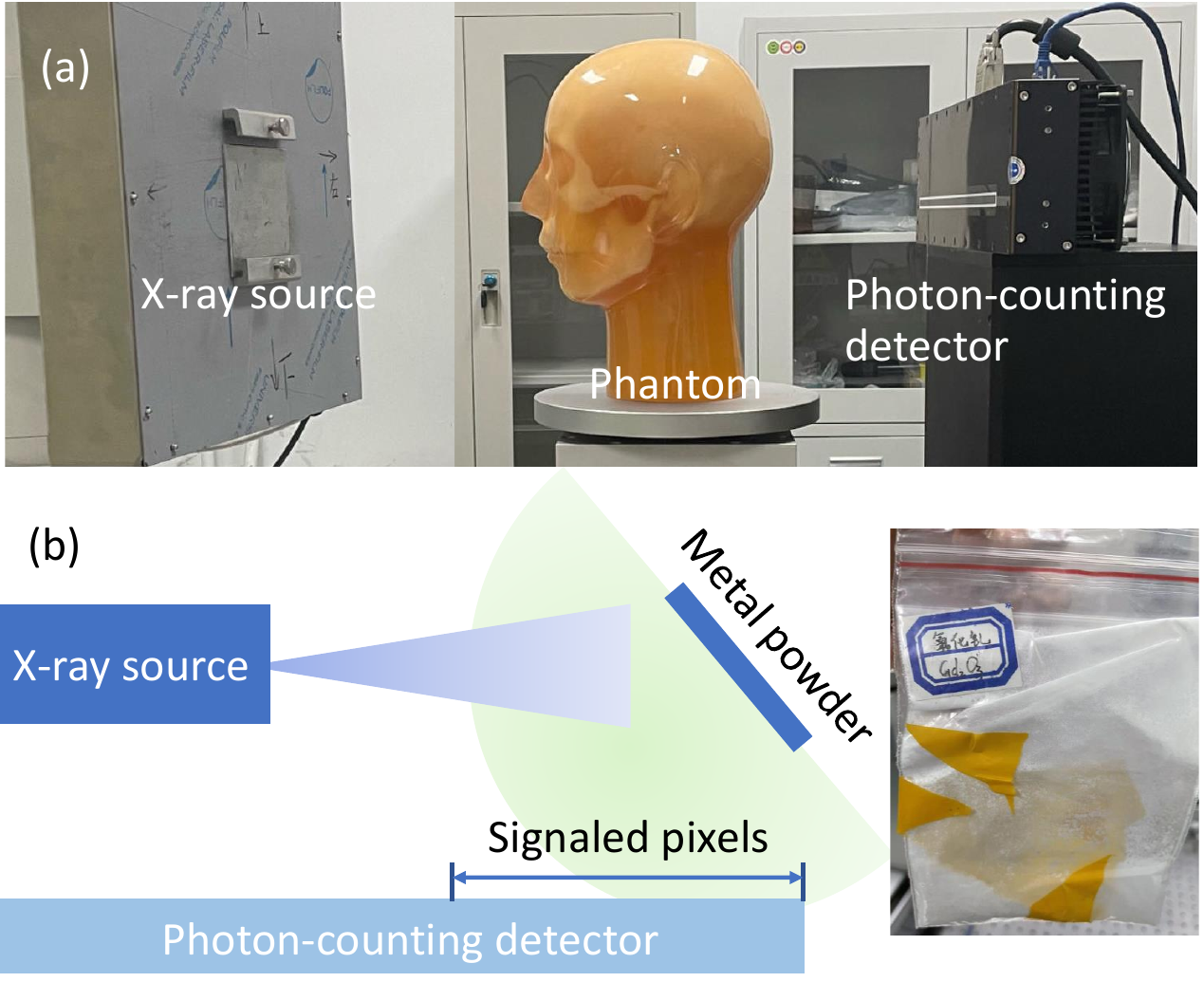}}
	\caption{Experiment setup of photon-counting CT: (a) Phantom experiment; (b) XRF-based calibration experiment. 
	}
	\label{fig:platform}
\end{figure}

The CT platform used for physical experiments is shown in Fig. \ref{fig:platform}(a). The Varex G242 tube with EMD high-voltage EPS50RF was adopted as the X-ray source in the system; it has a maximum voltage of 150 kV, a maximum power of 45 kW, and a focal size of 0.4 mm to 0.8 mm. A 2 mm aluminum filter was placed between the X-ray tube and the scanned object to harden the spectrum. The PCD employed in this study is XC-HYDRA FX35\cite{FX35}, with a CdTe sensor layer and two adjustable energy thresholds. The detailed imaging parameters of physical experiments are listed Table \ref{table:parameter}.

In our study, the source spectrum $S_{0i}$ was derived through expectation-maximization (EM) optimization\cite{sidky2005robust} of calibration data acquired from the a-Si Varex 4343 flat-panel detector within our experimental platform\cite{wang2023dual}. 

And the energy response function was simulated using GEANT4 for a 0.75 mm CdTe detector with 3584 × 64 pixels (0.1 mm $\times$ 0.1 mm pixel size). Charge sharing was modeled through a depth-dependent Gaussian charge cloud distribution,
where the charge distribution among neighboring pixels was calculated by integrating a 2D Gaussian function over a 3×3 pixel neighborhood\cite{broennimann2006pilatus,abramowitz1972handbook}, accounting for both the intrinsic charge spread at the detector surface and the depth-dependent charge cloud broadening. The $\bar{R}(E,E^\prime)$ with E = 40, 80, and 120 keV is shown in Fig. \ref{fig:energy response}.

Image reconstruction was all performed by the Filtered Backprojection (FBP) algorithm\cite{hsieh2022computed} with a hamming window in this work.

\begin{table}[!t]
	\caption{Parameters in physical experiments}
	\setlength{\tabcolsep}{1pt}
	\resizebox{255pt}{!}{
		\begin{tabular}{l@{\hspace{12pt}}l}
			\specialrule{1pt}{0pt}{0pt}
			\textbf{Parameter} & \textbf{Value} \\
			\hline
			Detector pixel size & 0.1 mm $\times$ 0.1 mm \\
			Sensor layer & 0.75 mm CdTe \\
			Detector pixel number & 3564 $\times$ 64 \\
			SID & 750 mm \\
			SDD & 1085 mm \\
			Frame rate & 20fps \\
			Acquisition frames & 1080 frames / 360° \\
			\hline
			\textbf{Scan} & \textbf{Voltage, Current, Energy threshold, AC mode}\\
			MEPT: Scan set-I & 120 kVp, 8.0 mA, 40/70 keV, AC OFF \\
			MEPT: Scan set-II & 120 kVp, 4.0 mA, 31/37/54/60 keV, AC OFF \\
			MEPT: Scan set-III & 120 kVp, 1.4/2.8/4.0/5.6 mA, 25/45 keV, AC ON \\
			MEPT: Scan set-MD & 120 kVp, 1.6 mA, 31/54 keV, AC ON \\ 
			HPhan: Scan set-I & 120 kVp, 3.2 mA, 40/60 keV, AC OFF\\ 
			HPhan: Scan set-II  & 080 kVp, 8.0 mA, 30/50 keV, AC OFF\\ 
			HPhan: Scan set-III & 120 kVp, 3.2 mA, 25/55 keV, AC ON\\ 
			HPhan: Scan set-IV & 080 kVp, 8.0 mA, 25/55 keV, AC ON\\ 
			HPhan: Scan set-MD & 120 kVp, 3.2 mA, 40/60 keV, AC OFF\\ 
			\hline

			Step wedge calibration & Al: 0, 0.94, 2.96, 5.04, 10.53, 20.67 (mm) \\
			Recon size & 512 $\times$ 512 \\
			Voxel size & 0.5 mm $\times$ 0.5 mm $\times$ 0.5 mm \\
			\specialrule{1pt}{0pt}{0pt}
			\multicolumn{2}{l}{\makecell[l]{Abbreviations: SID, Source to iso-center distance; SDD, Source to detector \\distance; MEPT, multi-energy phantom; HPhan, Head phantom.}}
		\end{tabular}
	}
	\label{table:parameter}
\end{table}

\begin{table}[!t]
	\caption{Parameters in numerical simulations}
	\setlength{\tabcolsep}{3pt}
	\resizebox{255pt}{!}{
			\begin{tabular}{l@{\hspace{12pt}}l}
			\specialrule{1pt}{8pt}{0pt}
			\textbf{Parameter} & \textbf{Value} \\
			\hline
			\multicolumn{2}{l}{\textbf{Validation I: Single-material Spectral Skew Term}} \\
			Introduction of uncorrected $g_i(E)$ & CdTe, $\{\Delta D_i\}$: -0.075 $\thicksim$ 0.1 mm \\
			Basis material in $g_i(E)$ calculation & CdTe \\
			Energy threshold setting& 40 / 60 keV\\
			$\Delta E_i$ introduced at $E_T$ = 60 keV & -10 $\thicksim$ 10 keV\\
			\hline
			\multicolumn{2}{l}{\textbf{Validation II: Dual-material Spectral Skew Term}} \\
			\multirow{6}{*}{Introduction of uncorrected $g_i(E)$} & CdTe, $\{\Delta L_{i}\}$: -0.02 $\thicksim$ 0.02 mm \\
			& Cu, $\{2\Delta L_i\}$: -0.04 $\thicksim$ 0.04 mm \\
			& Bone, $\{100\Delta L_i\}$: -2 $\thicksim$ 2 mm \\ 
			& Al, $\{50\Delta L_i\}$: -1 $\thicksim$ 1 mm \\ 
			& PMMA, $\{500\Delta L_i\}$: -10 $\thicksim$ 10 mm 
			\\
			Basis material in $g_i(E)$ calculation & CdTe \& Al \\
			Energy threshold setting & 40 / 60 keV\\
			$\Delta E$ introduced at $E_T$ = 60 keV & $\{\Delta E_{\mathrm{ref},i}\}$, -2 $\thicksim$ 2 keV\\
			\hline
			\multicolumn{2}{l}{\textbf{Simulated Abdominal Phantom}} \\
			Introduction of uncorrected $g_i(E)$ & CdTe, $\{3\Delta L_i\}$: -0.06 $\thicksim$ 0.06 mm \\
			Basis material in $g_i(E)$ calculation & CdTe \& Al \\
			Energy threshold setting& 40 / 60 keV\\
			$\Delta E$ introduced at $E_T$ = 60 keV & $\{0.5\Delta E_{\mathrm{ref},i}\}$, -1 $\thicksim$ 1 keV\\
			\specialrule{1pt}{0pt}{0pt}

		\end{tabular}
	}
	\label{table:simu parameter}
\end{table}

\subsection{Numerical Simulations}\label{Numerical Simulations}
To validate the proposed spectral model and ETB-Cal algorithm, we assessed accuracy and robustness of proposed model-based parameter-solving and evaluated reconstructed image quality of a simulated abdominal phantom. The detailed parameters in simulations are listed Table \ref{table:simu parameter}, with identical geometric parameters employed in physical experiments.

We conducted numerical simulations using a MATLAB-based custom-developed CT simulation tool that models X-ray emission, transmission, attenuation, and energy deposition in the detector. The forward projection was performed using the open-source TIGER toolbox functions\cite{biguri2016tigre}. This calculation determined the ray path length through the phantom for each detector pixel, enabling precise computation of X-ray attenuation ratios. These ratios were then incorporated into Eq.~\eqref{eq:TTF-spectralModel} to simulate the photon absorption process and subsequent signal generation in the detector.

We incorporated spectral inconsistency into the model through direct parameterization of CdTe thickness variations (\( \Delta D_{i} \)) and energy threshold bias (\( \Delta E_{i} \)). By introducing these non-uniformities into Eq.~\eqref{eq:TTF-spectralModel}, we simulated phantom scan counts and projection data containing spectral inconsistencies.

In addition, we performed five transmission measurements ($M = 5$): one air scan without any object in the beam, and four additional static scans of aluminum step wedges with thicknesses of 5 mm, 10 mm, 15 mm, and 20 mm, respectively. This measurement scheme allowed us to characterize the spectral inconsistency introduced in our numerical simulations.
\subsubsection{Validation of Spectral Skew Term and ETB Calculation}\label{Validation of Spectral Skew Term and ETB Calculation}

Using step wedge calibration data with 5 Al filters, we performed Validation I firstly (in Table \ref{table:simu parameter}), in which the spectral skew term was controlled by $\Delta D_{i}$ in both inconsistency generation and correction, and the inconsistency parameters were varied over an extended range to assess their impact on image artifacts. The results informed parameter selection and demonstrate the validity of ETB‑Cal.
As shown in Fig. \ref{fig:Simu validation}(1a), the distribution of $\Delta D_{i}$ consisted of a spatial high-frequency inconsistencies superimposed on a global low-frequency component. $g_{i}(E)$ was given by ${\rm e}^{-\mu_{\rm{1}}(E)\Delta D_{i}}$. $\Delta E_i$ followed a truncated Gaussian distribution over the interval [-10, 10] keV.

Furthermore, in Validation II (summarized in Table \ref{table:simu parameter}), CdTe and Al served as $\Delta D_{i}$ and $\Delta T_{i}$ in $g_{i}(E)$ during inconsistency estimation, and we applied refined parameter bounds of $\pm 0.02$ mm for $\Delta D_{i}$ and $\pm 2$ keV for $\Delta E_{i}$ when introducing inconsistencies (with the distributions shown in Fig. \ref{fig:Simu validation multi-mat}(a)). The $\Delta D_{i}$ and $\Delta E_{i}$ were hereafter denoted $\{\Delta L_i\}$ and $\{ \Delta E_{\mathrm{ref},i}\}$ in the subsequent sections.

To evaluate the robustness of spectral skew term calculation, we employed four additional materials (Cu, Al, bone, and PMMA) for spectral skew term introduction. The thickness inconsistencies were generated by scaling $\{\Delta L_i\}$ with factors of 2, 100, 50, and 500, respectively. All simulations maintained CdTe and Al as $\Delta D_{i}$ and $\Delta T_{i}$ in $g_{i}(E)$ for spectral skew term calculation, enabling comparative fidelity analysis across different materials. $\Delta E_{i}$ was constrained to the high-energy (HE) range [60, 120] keV to isolate its effects on $\Delta D_{i}$ and $\Delta T_{i}$ calculation.

\subsubsection{Simulated Abdominal Phantom}\label{Simulated Abdominal Phantom}

As summarized in Table \ref{table:simu parameter}, we introduced \{$\Delta D_{i}, \Delta E_{i}$\} = $\{3\Delta L_i, 0.5\Delta E_{\mathrm{ref},i}\}$ into the simulated abdominal phantom data to induce inconsistencies. We then used this dataset to evaluate the spectral inconsistency correction performance of ETB‑Cal and other methods (direct inversion, ring removal and two existing model-based methods, the implementations of latter two were displayed in Section \ref{Implementation of Two Existing Model-based Methods for Comparison} as mentioned in \ref{Some Details of Comparative Study}). We performed calibration within ETB‑Cal and direct inversion using the Al transmission data of five different thickness described formerly.

\subsection{Physical Experiments}

The phantoms used in the physical experiment were an anthropomorphic head phantom (Kyoto Kagaku, Kyoto, Japan) and a Gammex multi-energy phantom (Sun Nuclear Corp., Middleton, WI). The calibration measurements were carried out using the step wedges listed in Table \ref{table:parameter}. 

Data pre-processing in our work consisted of 2D spatial binning in two steps: 4-pixel horizontal averaging to mitigate charge-sharing effects and 10-row vertical averaging for noise suppression. Consequently, the photon counts from a single frame formed an 896 $\times$ 1 vector.

Experimental measurements suffer from pulse pileup, charge sharing, and scatter effects (especially in large phantoms). The detector features an anti-charge sharing (AC) functionality that reassigns neighboring pixel energies to the central pixel, thus suppressing charge sharing. When AC is enabled, pileup severity rises due to prolonged charge collection time.
\cite{ullberg2018photon,marupudi2025evaluation}. Conversely, AC disablement (AC-OFF) leads to a significantly enhanced charge sharing effect. 

Given the trade-off between pulse pileup and charge sharing, AC-OFF was used as the default setting in our work for two reasons: (1) spatial binning partially suppresses charge sharing, and (2) although increasing the tube current improves the signal-to-noise ratio, it exacerbates pileup. When AC-ON was required, the tube current was set to the minimum feasible value to limit pileup.
\subsubsection{Spectral Inconsisteny Correction}
As summarized in Table \ref{table:parameter}, we acquired CT scans of multi-energy phantom (MEPT) and head phantom under different combinations of kVp, current, energy threshold, and AC modes. Especially, in MEPT: Scan Set-II we assessed multi-threshold performance using two dual-threshold acquisitions (31/54 keV and 37/60 keV), generating spectral data at four distinct energy thresholds for ETB-Cal analysis. 

Moreover, given that pileup is common in PCCT and can induce non‑negligible spectral effects under certain conditions, we conducted complementary experiments to assess the compatibility between pileup correction and the ETB‑Cal method (see Appendix \ref{Pileup Correction Compatibility}) with AC enabled across varying tube currents (MEPT: Scan set-III in Table \ref{table:parameter}).

\subsubsection{Material Decomposition}
In our material decomposition experiments on MEPT containing three iodine rods (2, 5, 10 mg/mL) and a calcium rod embedded (MEPT: Scan set-III in Table \ref{table:parameter}), we selected AC ON mode with a reduced tube current of 1.6 mA to maintain quantitative accuracy by reducing spectral skew from charge sharing and suppressing pileup effects. 

However, in head phantom experiment, we conducted the experiment with AC OFF (Hphan: Scan set-V in Table \ref{table:parameter}) to reduce the influence of the pileup effect. Because we had to use higher tube current due to the strong attenuation of bone, which led to an insufficient number of X-ray photons reaching the detector, causing increased noise and severe streaky artifacts.

\subsection{Implementation of Two Existing Model-based Methods for Comparison}\label{Implementation of Two Existing Model-based Methods for Comparison}
Both methods (SFR and BCT) were implemented and evaluated in simulations and physical experiments. We used the same simulated abdominal phantom (with built-in spectral inconsistency) and the MEPT (Scan set‑I) to assess their performance and compare them against ETB‑Cal.
\subsubsection{SFR Model Method}

In the simulation studies, we constructed the 5 $\times$ 5 calibration data used in \cite{sidky2022spekcali} and implemented the calibration process proposed therein. This calibration dataset consisted of PMMA (thicknesses: 0, 2.5, 5, 7.5, 10 cm) combined with aluminum (thicknesses: 0, 0.6, 1.2, 1.8, 2.4 cm). In other words, the SFR method made use of more calibration data than ETB‑Cal and direct inversion.

Because of constraints of the experimental platform, we adapted the calibration protocol for the physical experiments. The original method recommends Al/PMMA filtration combinations, but we performed only separate measurements: six Al filters (summarized in Table \ref{table:parameter}, same as ETB-Cal) and eight PMMA slabs (18.7, 29.0, 37.8, 47.7, 76.5, 93.7, 107.5, 127.5 mm). This yielded 14 discrete calibration measurements without material combinations, resulting in fewer calibration data than those reported in \cite{sidky2022spekcali}.

\subsubsection{BCT Model Method}
The BCT model method was implemented through both simulation and physical experiments, implemented according to \cite{lee2018spekdisto}. This method requires creating a distinct Bias Correction Table (BCT) for each detector pixel during calibration to compensate for pixel-specific variations in the measurement vector $y_j$. In numerical simulations, this approach was evaluated with the artificial abdominal phantoms described in Section \ref{Simulated Abdominal Phantom}. For the physical experiment, data collection was restricted to only 37 sampling points, a number significantly lower than the desired value. This sparse sampling necessitated extensive interpolation and extrapolation during pixel-by-pixel BCT generation.

\subsection{Implementation of XRF-Based Threshold Bias Measurement}\label{Implementation of XRF-based Threshold Bias Estimation for Reference}
The X-ray fluorescence experiment was conducted under the setup shown in Fig. \ref{fig:platform}(b), with data processing of approximately 800 pixels that recorded signals of sufficient amplitude. Through Gaussian peak fitting of the count-versus-energy bin obtained by threshold sweeping flat-field scan (1 keV increment), we determined the empirical detector energy thresholds and quantified their deviations from known fluorescence monoenergies as ${\Delta E^k_{i}}$. The resulting threshold bias measurements were then compared with corresponding outputs from our ETB-Cal algorithm. To assess correction effectiveness, we performed image reconstruction using XRF-estimated parameters. The ${\Delta E^k_{i}}$ and ${g_{i}(E)}$ were extended from boundary of signaled detectors into adjacent regions via zero-order extrapolation.

\subsection{Evaluations}

To quantify the spectral inconsistency in reconstructed images, we measured the maximum difference of CT numbers among selected regions of interest (ROIs) of the object, i.e.
\begin{equation}
	\begin{aligned}
		{\rm {\Delta}_{MAX}} = {\rm max}(\mu _n) - {\rm min}(\mu _n),
	\end{aligned}
\label{eq:Inco} 
\end{equation}

\noindent where $\mu_n$ is the averaged value inside the $n$-th ROI. 
And the image noise level was measured by the standard deviation (STD) in this work.

Also, to quantitatively evaluate the discrepancies and similarities between reconstructed images $I$ and reference images $\hat{I}$ in numerical simulations, the Mean Squared Error (MSE) the Structural Similarity Index (SSIM) were employed as

\begin{equation}
	\mathrm{MSE}(I, \hat{I}) = \frac{1}{N}\sum_{i=1}^{N} (I_i - \hat{I}_i)^2
\end{equation}

\noindent and
\begin{equation}
	\mathrm{SSIM}(I,\hat{I}) = \frac{(2\mu_I\mu_{\hat{I}} + C_1)(2\sigma_{I\hat{I}} + C_2)}{(\mu_I^2 + \mu_{\hat{I}}^2 + C_1)(\sigma_I^2 + \sigma_{\hat{I}}^2 + C_2)}
\end{equation}
where $N$ is the total number of pixels in the image, $\mu$ denotes the mean value, $\sigma^2$ denotes the variances, $\sigma_{I\hat{I}}$ represents the covariance, $C_1$, $C_2$ are stabilization constants.

To evaluate ETB‑Cal comprehensively, we conducted assessments including: (1) the $\Delta_{\rm{MAX}}$ within selected ROIs in the MEPT and head phantom under several different scan protocols, (2) the average STD within selected ROIs, and (3) quantitative MD accuracy on the MEPT and qualitative MD performance on the head phantom.
\begin{figure}[!t]
	\centering
	\centerline{\includegraphics[width=\columnwidth]{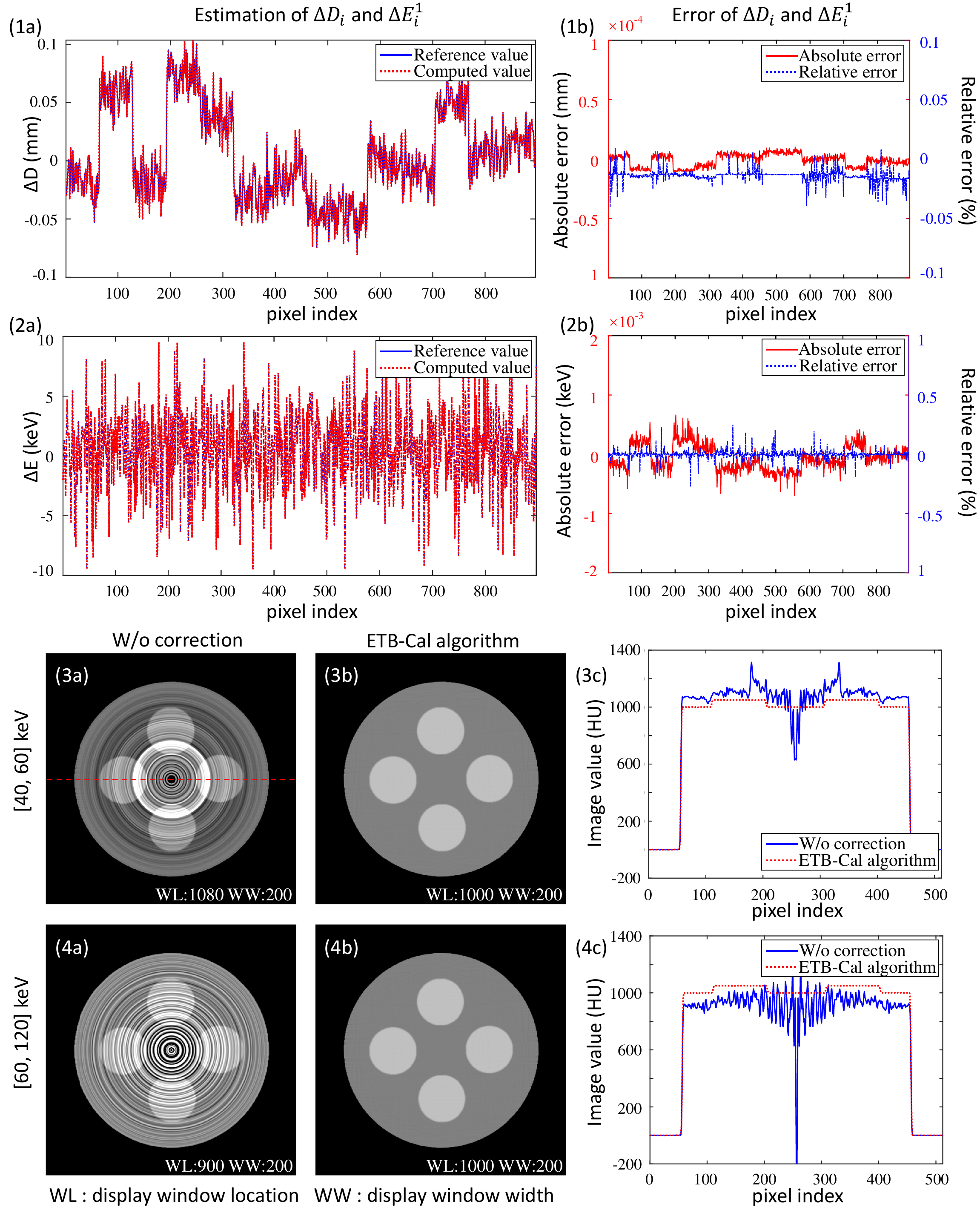}}
	\caption{{Parameter-solving accuracy of ETB-Cal algorithm with a single-material spectral skew term. The $\Delta D_{i}$ (selected as CdTe) distributed from -0.075 to 0.1 mm, which accounted for both inter-module and intra-module random variations, and the $\Delta E_{i}$ from -10 keV to 10 keV following a truncated gaussian distribution were incorporated. Both the absolute and relative errors in the solutions of $\Delta D_{i}$ and $\Delta E_{i}$ were minimal, and excellent results were achieved in correcting inconsistencies in the reconstructed images of a virtual water phantom containing four slightly denser (1.05 times) regions, without the addition of noise.}}
	\label{fig:Simu validation}
\end{figure}

For the datasets (simulated abdominal phantom and MEPT: Scan set-I) on which we compared our proposed method with four methods, we would present the results in two parts:

In the simulated abdominal phantom study, we compared MSE, SSIM and $\Delta_{\rm{MAX}}$ separately in Section \ref{results: Simulated Abdominal Phantom} (direct inversion and ring removal approaches) and Section \ref{results: Comparison with Two Existing Model-based Methods} (two existing model-based methods). In the MEPT study, we compared $\Delta_{\rm{MAX}}$ separately in Section \ref{Gammex Multi-Energy Phantom} and Section \ref{results: Comparison with Two Existing Model-based Methods}.

In particular, for SFR, which performed comparably to ETB-Cal in the simulation study, we also compared its MD performance with ETB-Cal on the MEPT and the head phantom in Section \ref{Comparison with SFR Model Method}.

\begin{figure*}[!t]
	\centering
	\centerline{\includegraphics[width=1.9\columnwidth]{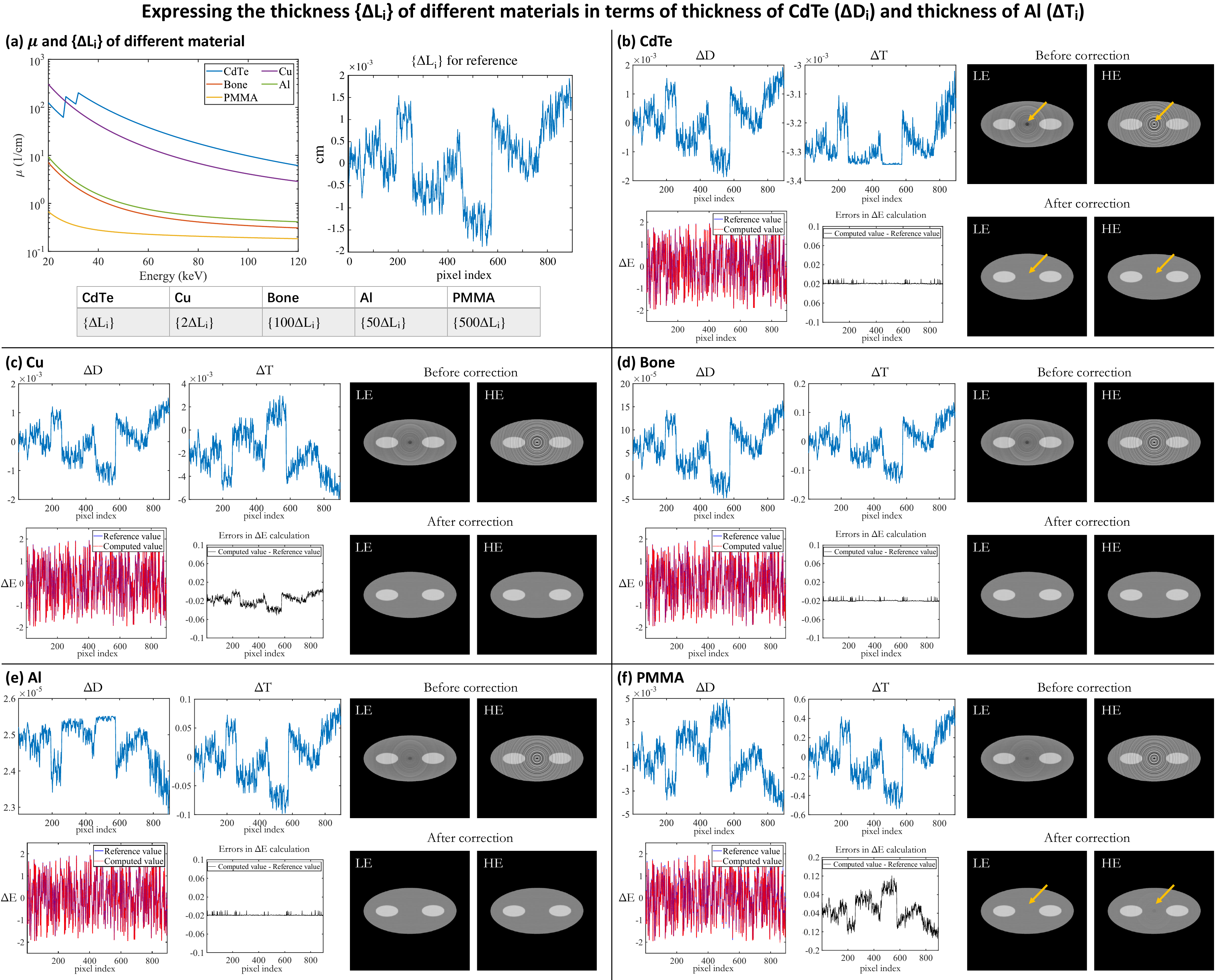}}
	\caption{{Parameter-solving accuracy of ETB-Cal algorithm with a dual-material spectral skew term. (a): Five materials selected as the inconsistency-inducing materials. Thickness inconsistency of these materials were derived by scaling the reference value $\{\Delta L\}$ with a constant factor. (b)-(f): The calculated $\Delta D_i$, $\Delta T_i$, and $\Delta E_i$ with CdTe and Al as the basis materials for the spectral skew term, together with reconstructed images of the virtual phantom before and after correction.
	}}
	\label{fig:Simu validation multi-mat}
\end{figure*}

\begin{figure}[!t]
	\centering
	\centerline{\includegraphics[width=1\columnwidth]{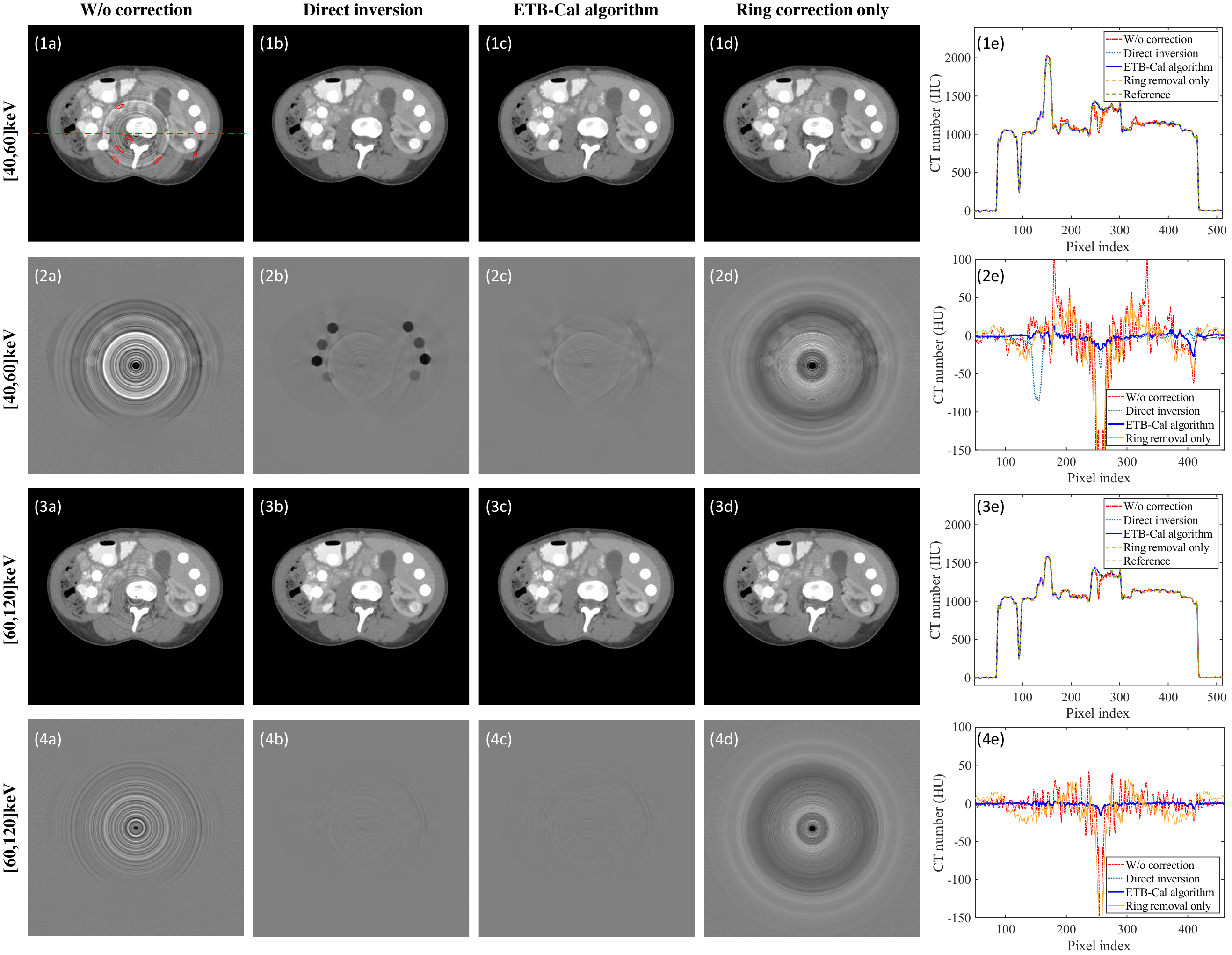}}
	\caption{Reconstructed images of simulated abdominal phantom: (a) with global bhc (a conventional bhc with no pixel-by-pixel spectral inconsistency compensation) and no any additional correction; (b) with direct inversion; (c) with the proposed ETB-Cal algorithm; (d) with a ring removal only. The profile line illustrates the CT values of the abdominal phantom along the red line shown in (1a). In this numerical simulation, the geometric parameters were consistent with physical experiment, with no noise introduced. Display window, for simulated abdominal phantom: [800, 1300] HU, for diff images: [-100, 100] HU.}
	\label{fig:simu result}
\end{figure}

\section{Results}\label{Results}

\subsection{Numerical Simulations}

\subsubsection{Accuracy of ETB-Cal Algorithm}
Parameter-solving accuracy of ETB-Cal algorithm with a single-material spectral skew term was shown in Fig. \ref{fig:Simu validation}. The reconstructed images of the virtual water phantom and their corresponding centerline profiles demonstrated complete elimination of severe artifacts caused by predefined spectral inconsistency. Quantitative analysis demonstrated that the absolute error of $\Delta D_{i}$ remained within \( 2 \times 10^{-5} \) mm (relative error $<0.05\%$), while $\Delta E_{i}$ exhibited absolute errors below \( 1 \times 10^{-3} \) keV (relative error $<0.3\%$), 
Some outliers of calculated $\Delta D_{i}$ and $\Delta E_{i}^1$ occurred in relative errors when their true values approached zero.
Parameter-solving accuracy of ETB-Cal algorithm with a dual-material spectral skew term was shown in Fig. \ref{fig:Simu validation multi-mat}. 
The artifacts in the simulated phantom were effectively mitigated when using CdTe and Al as the basis material for $g_i(E)$ across various inconsistency introducing materials. The errors in $\Delta E_i$ were under $0.05$ keV in most situations. However, degraded performance occurred with PMMA due to significant discrepancies between its attenuation properties and those of CdTe and Al. These results demonstrated the overall effectiveness of our method despite is material-dependent in theory.

\subsubsection{Simulated Abdominal Phantom}\label{results: Simulated Abdominal Phantom}

Reconstructed images of simulated abdominal phantom and its centerline profiles were shown in Fig. \ref{fig:simu result} (see the results of two existing model-based methods in Fig. \ref{fig:comparison}). ETB-Cal effectively mitigated ring artifacts in the simulated abdominal phantom, as shown in panels (1c)-(4c), whereas the ring removal approach retained significant residual artifacts and direct inversion method output inaccurate values in the regions containing iodine. In addition, the line profile in (2e) and (4e) revealed that ETB-Cal achieved smallest diviations from the reference images. 

We estimated the MSE and SSIM between the reconstructed images and the reference images of the simulated abdominal phantom for all non-air regions. Additionally, we compared $\Delta_{\rm{MAX}}$ on the E-bin difference images among the selected ROIs marked in Fig. \ref{fig:simu result}(1a). MSE, SSIM, $\Delta_{\rm{MAX}}$ in ETB-Cal corrected images and the difference images were \( 4.45 \times 10^{-7} \), 0.9999, 3.0 Hounsfield unit (HU) in LE image and \( 7.64 \times 10^{-8} \), 1.0000, 0.8 HU in HE image, respectively. While the values achieved by direct inversion method were \( 6.39 \times 10^{-6} \), 0.9999, 3.6 HU in LE image and \( 7.73 \times 10^{-8} \), 1.0000, 1.0 HU in HE image. The results achieved by ring removal approach were markedly worse.

\subsection{Physical Experiments}\label{results: Physical Experiments}

\subsubsection{Gammex Multi-Energy Phantom}\label{Gammex Multi-Energy Phantom}

\begin{figure*}[!t]
	\centering
	\centerline{\includegraphics[width=2\columnwidth]{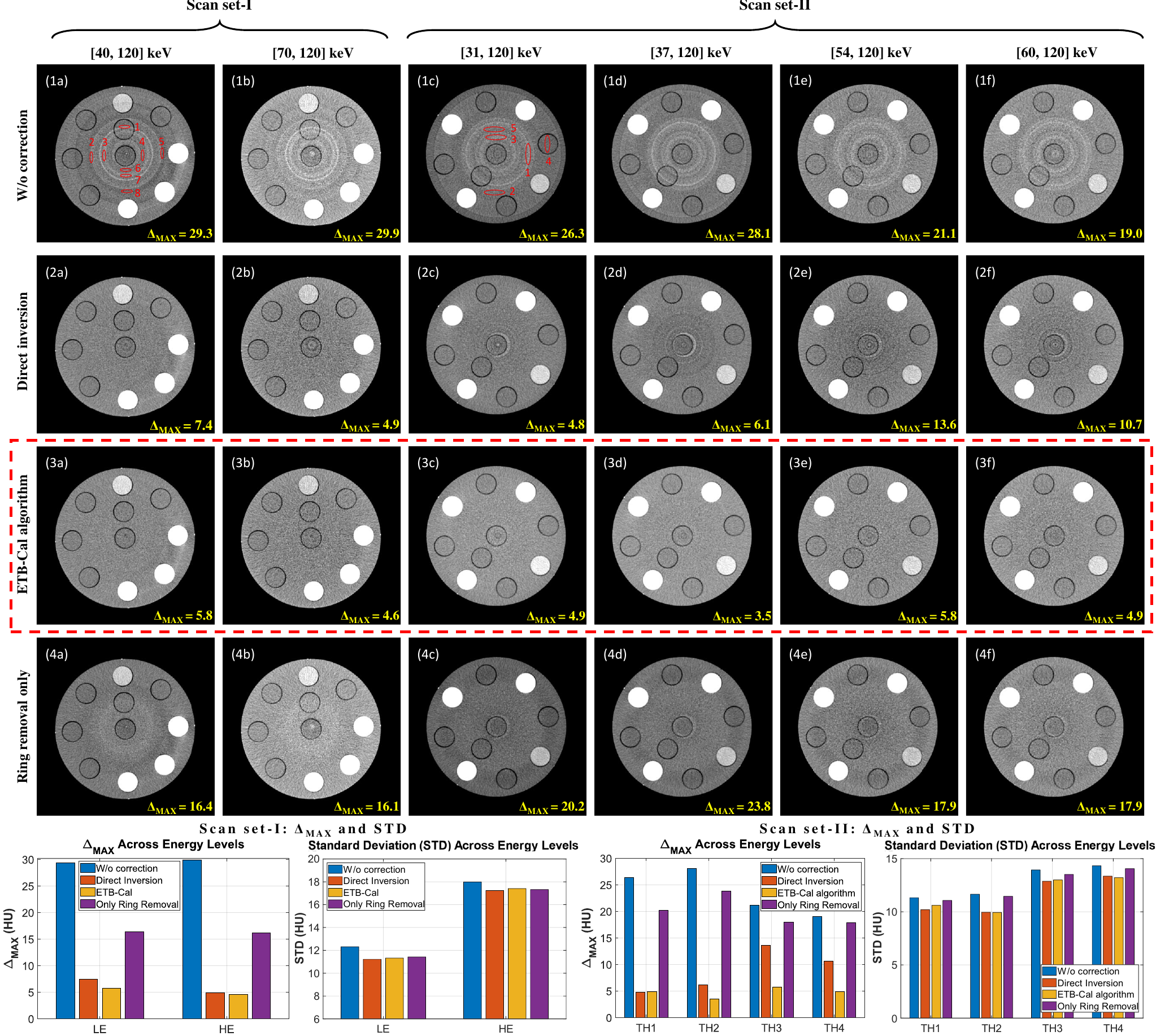}}
	\caption{Reconstructed images of Gammex multi-energy phantom under AC OFF mode, Scan set-I: 120 kVp, 8 mA, 20fps, 40/70 keV energy threshold, Scan set-II: 120 kVp, 4 mA, 20fps, 31/37/54/60 keV energy threshold: (1) with global bhc (a conventional bhc with no pixel-by-pixel spectral inconsistency compensation) and no any other calibration; (2) with direct inversion calibration; (3) with ETB-Cal algorithm calibration; (4) with only the ring removal method calibration. Total of 8 selected ROIs in Scan set–I phantom were marked in (1a). Total of 5 selected ROIs in Scan set–II phantom were marked in (1c). For ETB-Cal algorithm, photon counts in [40, 120] keV E-bin of Scan set-I and [31, 120] keV E-bin of Scan set-II were selected to obtain the $g_{i}(E)$, making only $\Delta E^k_{i}$ (k = 1 in Scan set-I and k = 1, 2, 3 in Scan set-II) need to be calculated in other E-bins. Display window: [900, 1100] HU.}
	\label{fig:ETB-cal at different TH}
\end{figure*}

Reconstructed images of Gammex MEPT was shown in Fig. \ref{fig:ETB-cal at different TH}.
Eight ROIs were selected across varying intensity levels of the ring artifacts, as shown in (1a), to assess artifact correction efficacy by calculating \( \Delta_{\rm{MAX}} \) in Scan set-I. For the [40, 120] keV energy bin (E-bin), the \( \Delta_{\rm{MAX}} \) values after correction with the direct inversion method, ETB-Cal algorithm, and ring removal method decreased from 29.3 HU to 7.4 HU, 5.8 HU and 16.4 HU, respectively. Similarly, for the [70, 120] keV E-bin, the \( \Delta_{\rm{MAX}} \) values after correction with the direct inversion method, ETB-Cal algorithm, and ring removal method decreased from 29.9 HU to 4.9 HU, 4.6 HU and 16.1 HU, respectively. And as marked in (1c), five ROIs were selected for Scan set-II to compute \( \Delta_{\rm{MAX}} \). The ETB-Cal algorithm achieved significant ring artifact reduction across all E-bins, with \( \Delta_{\rm{MAX}} \) reductions from 26.3 HU to 4.9 HU, from 28.1 HU to 3.5 HU, from 21.1 HU to 5.8 HU, and from 19.0 HU to 4.9 HU in four E-bins, respectively.

Additionally, we calculated the standard deviation (STD) within the two sets of selected ROIs and used the average STD to characterize noise levels. As shown in the bottom panel of Fig. \ref{fig:ETB-cal at different TH}, the noise levels in the corrected images with ETB-Cal were comparable to those in images with only global BHC (i.e., labeled 'w/o correction' in the figure). Moreover, after removing some outliers associated with ring artifacts in the reconstructed images, the average STD decreased slightly by 0.7-1.2 HU across different threshold settings.

\subsubsection{Kyoto Head Phantom}\label{Kyoto Head Phantom}

\begin{figure}[!t]
	
	\centering
	\centerline{\includegraphics[width=1\columnwidth]{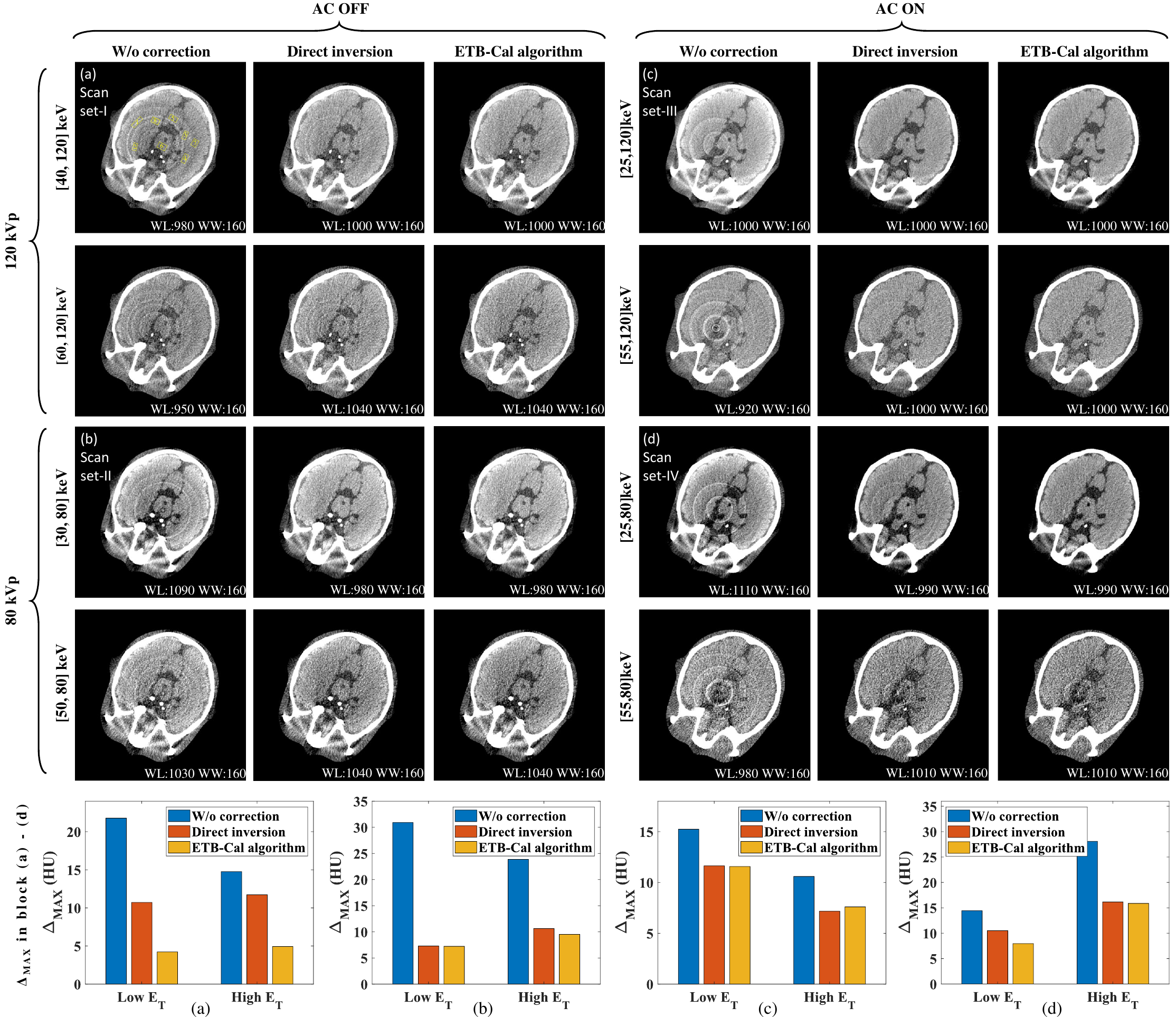}}
	\caption{Reconstructed images of Kyoto Head phantom under 120 kVp, 3.2 mA / 80 kVp, 8 mA and four different threshold set before and after calibration: (a) Scan set-I: 120 kVp, 3.2 mA, 40/60 keV energy threshold, AC OFF (b) Scan set-II: 80 kVp, 8 mA, 30/50 keV energy threshold, AC OFF (c) Scan set-III: 120 kVp, 3.2 mA, 25/55 keV energy threshold, AC ON (d) Scan set-IV: 80 kVp, 8 mA, 25/55 keV energy threshold, AC ON. The values of inconsistency ($\Delta_{\rm{MAX}}$) for the Kyoto head phantom under different experimental conditions were displayed at the bottom.}
	\label{fig:head phantom}
\end{figure}

Fig. \ref{fig:head phantom} presents reconstructed images of the Kyoto Head phantom acquired under several scan protocols (summarized in Table \ref{table:parameter}). The images indicate that the ETB-Cal algorithm achieves improved ring-artifact suppression relative to direct inversion in most cases. However, with AC ON, the artifacts were not fully removed, which may be attributable to aggravated pileup effects. In addition, we performed $\Delta_{\rm{MAX}}$ quantitative analysis in selected 8 ROIs (marked with yellow lines in (a)), with the results shown at the bottom of Fig. \ref{fig:head phantom}. In terms of $\Delta_{\rm{MAX}}$, the ETB-Cal algorithm also outperforms direct inversion. In particular, for Scan set-I, images corrected with ETB-Cal had $\Delta_{\rm{MAX}}$ values that were 6.5 keV and 6.8 keV lower than those from direct inversion in the [40, 120] keV and [70, 120] keV E-bins, respectively.

It is observed that the direct inversion method underperformed our proposed ETB‑Cal as expected. This is because the direct inversion attempts to solve the spectral skew term and threshold bias term in Eq. \eqref{eq:TTF-spectralModel} directly, which makes the solution space more ill-conditioned, as the two terms have some equivalence in softening or hardening the spectrum, leading the solution to often fall into a suboptimal local minimum. Additionally, the linear approximation solving in Eqs. \eqref{eq:ETB linear approx} and \eqref{eq:giE linear approx} assumes $\Delta E_i^k$ and $\mu_1\Delta D_i{+\mu}_2\Delta T_i$ are small values (close to zero), making direct inversion less stable where such assumptions on $\Delta E_i^k$, and $\Delta D_i$ and $\Delta T_i$ need to be valid simultaneously, while ETB-Cal only requires them to be valid sequentially.
\subsection{Comparison with Two Existing Model-Based Methods}\label{results: Comparison with Two Existing Model-based Methods}

\subsubsection{Comparison with SFR Model Method}\label{Comparison with SFR Model Method}
\begin{figure}[!t]
	\centering
	\centerline{\includegraphics[width=1\columnwidth]{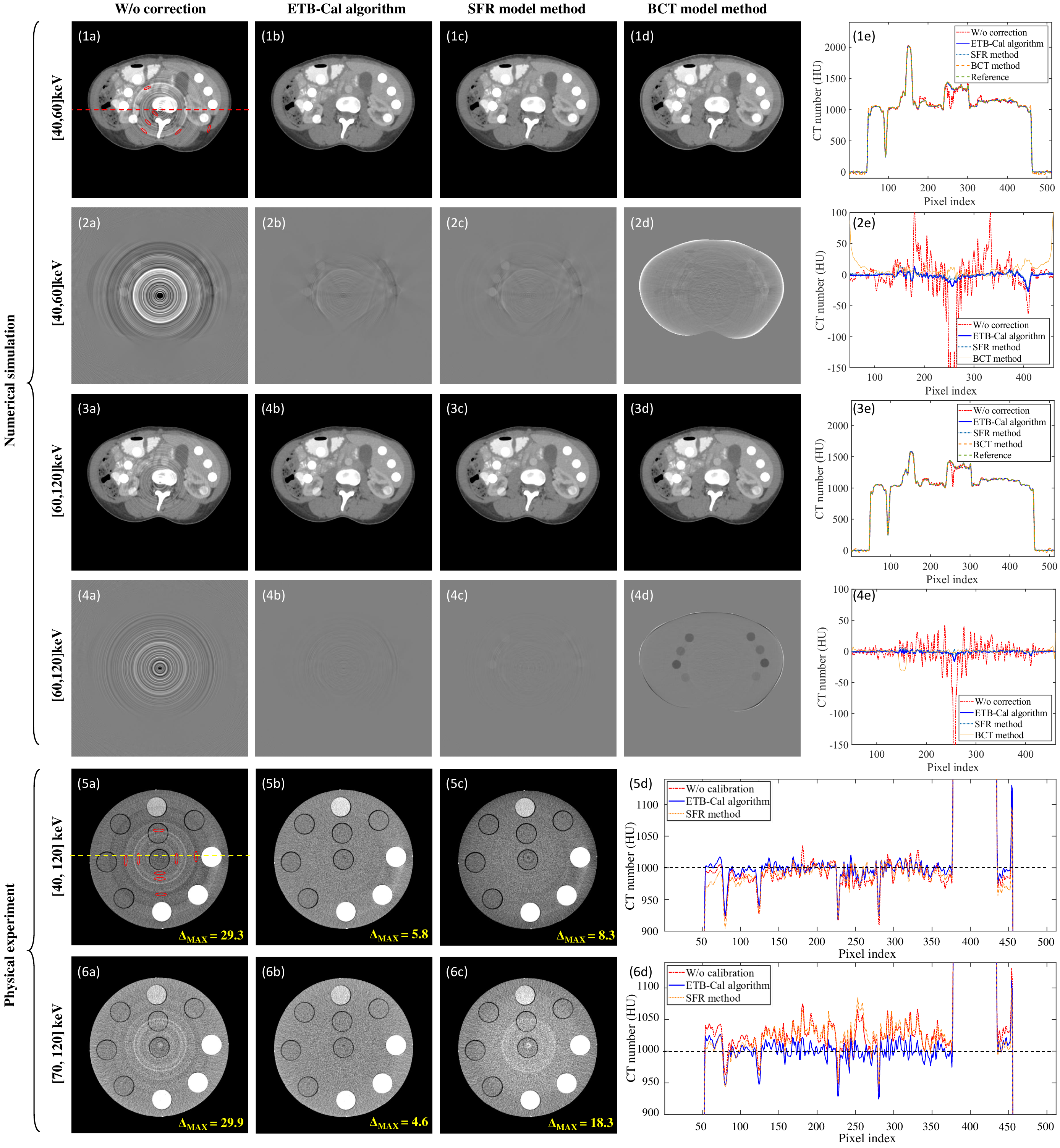}}
	\caption{Comparison with two existing model-based methods using reconstructed images of simulated abdominal phantom and multi-energy phantom: (a) with global bhc (a conventional bhc with no pixel-by-pixel spectral inconsistency compensation) and no any other calibration; (b) with ETB-Cal algorithm calibration; (c) with SFR model method; (1d)-(4d) with BCT model method. Row 2 and 4: reference-relative diff images of corresponding images in Row 1 and 3. Line profiles showed the CT values of the simulated abdominal phantom and the multi-energy phantom along the red and yellow lines, respectively, as marked in (1a) and (5a). Display window, for simulated abdominal phantom: [800, 1300] HU, for diff images: [-100, 100] HU, for Gammex multi-energy phantom: [900, 1100] HU.}
	\label{fig:comparison}
\end{figure}

In simulation studies, it can be seen in Fig. \ref{fig:comparison}(1a) that ETB-Cal, which utilized only five Al filters for calibration, showed comparable correction capability with SFR method which used 5 × 5 step wedges combinations.
And quantitative evaluation in Table \ref{table:simutable} demonstrated that the SFR method achieved MSE, SSIM, $\Delta_{\rm{MAX}}$ values of $3.66 \times 10^{-7}$, 0.9999, 3.3 HU and $5.22 \times 10^{-8}$, 1.0000, 0.7 HU for the reconstructed LE and HE images, respectively, which were slightly better than that achieved with ETB-Cal.

In the physical experiments, with 14 measurements of calibration data (while only 6 Al filters for ETB-Cal), both LE and HE images processed with the SFR method exhibited reduced quality. Without correction, $\Delta_{\mathrm{MAX}}$ was 29.3 HU ([40, 120] keV E-bin) and 26.3 HU ([70, 120] keV E-bin). After correction, the ETB-Cal algorithm reduced them to 5.8 HU and 4.6 HU, respectively, while the SFR method decreased them to 8.3 HU and 18.3 HU.

Firstly, the SFR method shows a comparable \( \Delta_{\rm{MAX}} \) statistic to ETB-Cal on the LE images. However, as shown in the profile line of Fig. \ref{fig:comparison}(5d), the reconstructed images of MEPT exhibit a general trend of higher HU values in the center and lower HU values at the edges. The distribution of HU values in the water region is clearly uneven. Moreover, the correction effect of SFR on the HE images is significantly inferior to that of ETB-Cal, with a higher \( \Delta_{\rm{MAX}} \). As seen in Fig. \ref{fig:comparison}(6c), the corrected images also show prominent ring artifacts. Additionally, the profile line in Fig. \ref{fig:comparison}(6d) reveals similar issues as those observed in the SFR-corrected LE images.

\begin{figure*}[!t]
	\centering
	\centerline{\includegraphics[width=2.1\columnwidth]{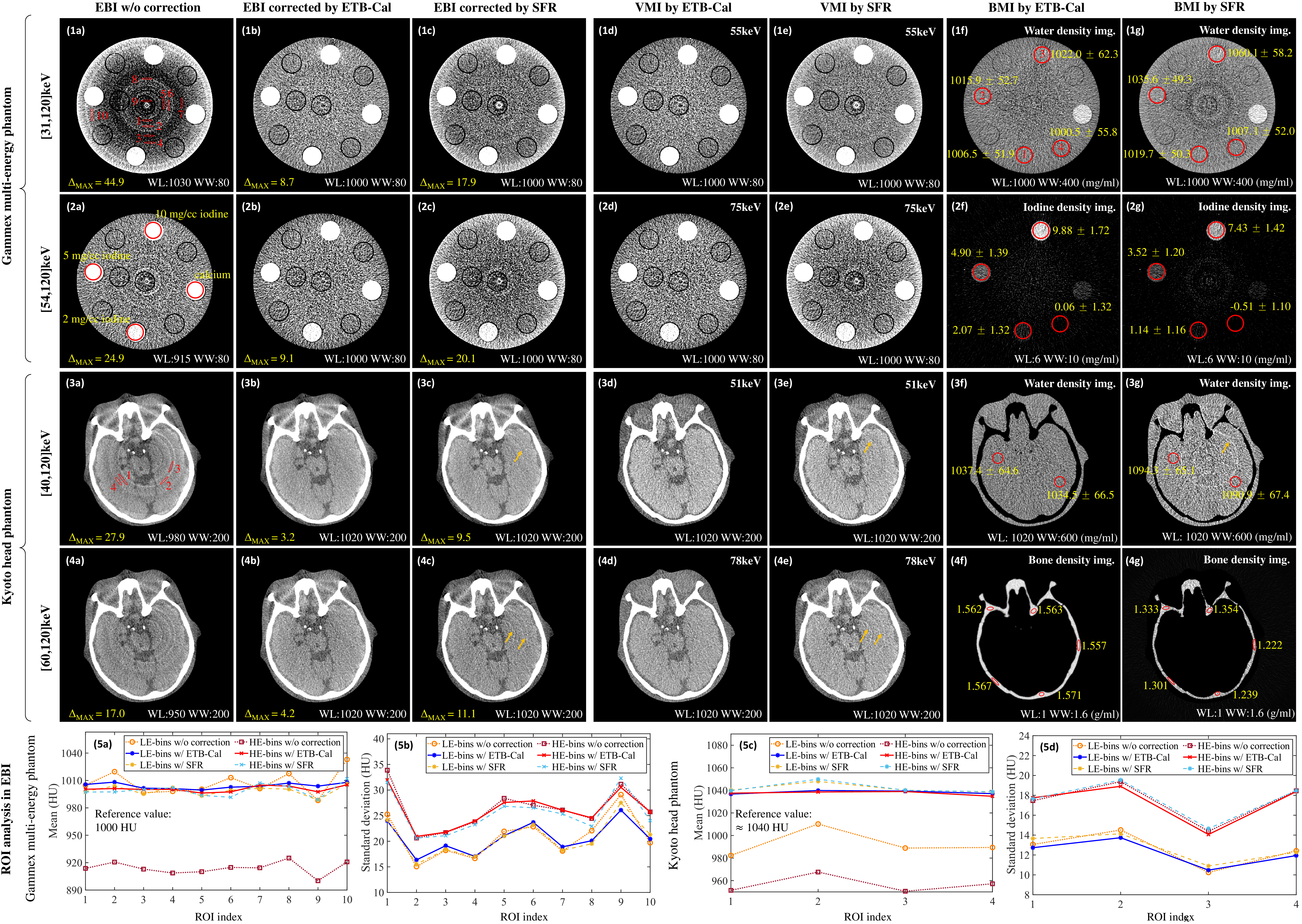}}
	\caption{Material decomposition results using the ETB-Cal and SFR methods for the Gammex multi-energy phantom (rows 1 \& 2; AC ON, 120 kVp, 1.6 mA, 31/54 keV threshold) and for Kyoto head phantom (rows 3 \& 4; AC OFF, 120 kVp, 3.2 mA and 40/60 keV energy threshold). The iodine concentrations within the three iodine rods are 2, 5, and 10 mg/ml. The ROIs have been marked on the reconstruction images. Abbreviations: EBI, E-bin image; VMI, Virtual mono-chromatic image; BMI, Basis material image. Unit in EBI and VMI: HU.}
	\label{fig:MD}
\end{figure*}

Material decomposition results of Gammex MEPT was shown in Fig. \ref{fig:MD}.
As marked in (2a), the reference concentrations of three iodine rods in MEPT were 2, 5, and 10 mg/ml. Meanwhile, the estimated iodine concentrations by ETB-Cal in (2f) were 2.07, 4.90, and 9.88 mg/ml. While the measured water densities in the selected ROIs in (1f) were 1006.5, 1015.9, and 1022.0 mg/ml, demonstrating slight deviations from the reference water density (1000.0 mg/ml). The estimated iodine concentrations by SFR in (2g) were 1.14, 3.52, and 7.43 mg/ml. While the measured water densities in the selected ROIs in (1g) were 1019.7, 1035.6, and 1060.1 mg/ml. SFR deviated more from the reference values than ETB-Cal.

It can be seen in Fig. \ref{fig:MD}(3f) and (4f) that bone and soft tissue of the head phantom were completely separated, and for VMIs at 51 keV and 78 keV, both the ring artifacts and the bony beam hardening artifacts were significantly reduced.
In (4f), We selected five ROIs in the bone region for mean value statistics. Although it is challenging to do MD in head phantom, the range of values relative to the smallest mean was only 0.9 \%, indicating that the decomposition values were fairy accurate and exhibited good uniformity. For SFR, water and bone are also separated in (3g) and (4g). However, both show some value offsets, and the water image still contains residual artifacts.

We also compared the \( \Delta_{\rm{MAX}} \) on the E-bin images before and after ETB-Cal correction among the selected ROIs marked in Fig. \ref{fig:MD}(1a) and (3a). 
For the MEPT, \( \Delta_{\rm{MAX}} \) decreased 44.9 HU to 8.7 and 17.9 HU in the [31, 120] keV E-bin, and from 24.9 HU to 9.1 and 20.1 HU in the [41, 120] keV E-bin, with ETB-Cal and SFR, respectively.
For the head phantom, \( \Delta_{\rm{MAX}} \) decreased from 27.9 to 3.2 and 9.5 HU in the [40, 120] keV E-bin, and from 17.0 to 4.2 and 11.1 HU in the [60, 120] keV E-bin, with ETB-Cal and SFR, respectively.
The individual mean values and standard deviations (STD) for each ROI were plotted in (5a)-(5d).

SFR performed worse than ETB-Cal because the calibration data did not form a two-dimensional grid. As a result, for the polynomial operation in Eq. \eqref{eq:Talpha}, the data points were not dense enough, which biased the polynomial correction and led to an overall bias in the reconstructed images. This was especially clear in Fig. \ref{fig:MD}(1c) and (2c).

\subsubsection{Comparison with BCT Model Method}
It can be seen in Fig. \ref{fig:comparison} that the BCT method performed worse than ETB-Cal in simulated abdominal phantom, and as illustrated in Table \ref{table:simutable}, the BCT method achieved MSE, SSIM, $\Delta_{\rm{MAX}}$ values of $1.80 \times 10^{-5}$, 0.9964, 5.8 HU and $2.03 \times 10^{-6}$, 0.9991, 1.6 HU for the reconstructed LE and HE images, respectively. This degradation occurred because accuracy of BCT method changes with material type and thickness (as reported in \cite{lee2018spekdisto}), making it harder to correct the added inconsistencies in our study.

The platform constraint led to severely insufficient number of sampling points for reliable BCT generation. This limitation forced extensive interpolation/extrapolation, introducing significant errors that undermined the experimental validity of BCT method. To avoid misinterpretation, the results in physical experiment were not included here.

Based on the preliminary results above, our implementation of the two existing model-based methods yielded satisfactory performance in simulations, yet exhibited limited efficacy in physical experiments. Comparing with the two existing model-based methods, our proposed ETB-Cal approach demonstrated relatively superior performance, providing effective spectral correction and artifact reduction with merely five Al filters.

It was noted that the SFR method demonstrated limited experimental efficacy due to two main factors: (1) failure to achieve the 25 Al/PMMA thickness combinations required by \cite{sidky2022spekcali}, combined with dependence on matching both material composition and thickness between calibration filters and objects to be corrected and (2) high-energy spectrum deviations impacting $\alpha$ and $\beta$. 
Our BCT implementation faced three principal constraints: (1) potentially insufficient iteration for parameter optimization, (2) inherent thickness-dependent estimation errors as reported in \cite{lee2018spekdisto}, and (3) stringent sampling requirements—specifically, the need for extensive and densely distributed calibration points (0.05 × 0.05 interval over 40 × 200 coverage points as specified in \cite{lee2018spekdisto}) that proved experimentally unattainable. These limitations restricted the reliable application to simulations where ideal sampling conditions and spectrum accuracy can be guaranteed.

\subsection{Comparison with XRF Method}
\begin{figure}
	\centering
	\centerline{\includegraphics[width=1\columnwidth]{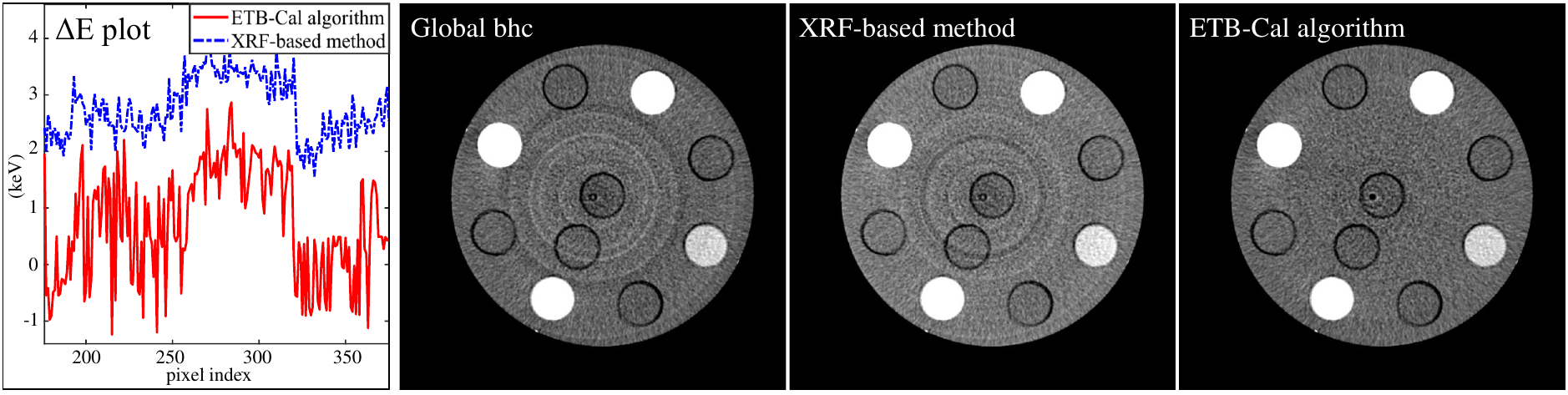}}
	\caption{Results of XRF experiment. The XRF-based correction extended the ${\Delta E^k_{i}}$ measured in the signaled area to the entire detector row and set ${g_{i}(E)}$ to 1. Display window: [920, 1080] HU.}
	\label{fig:XRF-Comp}
\end{figure}

Results of the XRF experiment were shown in Fig. \ref{fig:XRF-Comp} and indicated that the XRF‑based method performed suboptimally. This outcome was attributable to experimental constraints: only 800 of 3596 pixels exhibited XRF signal of sufficient intensity for reliable estimation, and an overly simplistic zero-order extrapolation was applied to the remaining 2796 pixels. In addition, we estimated threshold biases (${\Delta E^k_{i}}$) by comparing specific energies from the fluorescence emissions of metal powders with detector thresholds determined by a Gaussian-fitted energy-count curve from threshold sweeping scans. As such an experiment cannot account for the spectral skew, in Fig. \ref{fig:XRF-Comp} a global shift of about 2 keV is observed between the two threshold bias curves by the XRF-based and ETB-Cal approaches. Therefore, the XRF results can only serve as an approximate reference due to the experimental limitations in our work as described in Section \ref{Implementation of XRF-based Threshold Bias Estimation for Reference}. Nevertheless, the two curves showed similar trends as the threshold bias changes across pixels, thereby still confirming the reliability of the ${\Delta E^k_{i}}$ values calculated by the ETB-Cal.  

\begin{table}[t!]
	\centering
	\caption{Mean squared error and structural similarity between reference values and corrected images obtained by different methods and non-uniformity in diff images in Fig. \ref{fig:simu result} and simulated abdominal phantom in Fig. \ref{fig:comparison}.}
	\begin{tabular}{ccccccc}
		\specialrule{1pt}{0pt}{0pt}
		\diagbox{}{} & \multicolumn{2}{c}{\textbf{MSE}} & \multicolumn{2}{c}{\textbf{SSIM}}  & \multicolumn{2}{c}{$\boldsymbol{\Delta_{\mathrm{MAX}}}$} \\ \hline
		\textbf{Imgs. with} & LE & HE & LE & HE & LE & HE \\ \hline
		w/o correction & 2.70e-5 &	5.78e-6    & 0.9917&	0.9974 & 14.2 & 10.5  \\ 
		Direct inv. &6.39e-6&	7.73e-8  &0.9982&	1.0000 & 3.6 & 1.0\\ 
		\textbf{ETB-Cal alg.}  & 4.45e-7&	7.64e-8    &0.9999&	1.0000 & 3.0 & 0.8  \\ 
		RRO &1.59e-5&	9.52e-6 &0.9952&	0.9961  & 14.9 & 17.9 \\ 
		SFR Meth.   & 3.66e-7&	5.52e-8    &0.9999&	1.0000   & 3.3 & 0.7 \\ 
		BCT Meth.  &1.80e-5&	2.03e-6&0.9964&	0.9991 & 5.8 & 1.6  \\ 
		\specialrule{1pt}{0pt}{0pt}
		\multicolumn{7}{l}{\makecell[l]{Abbreviations: inv. for inversion; alg. for algorithm; RRO for Ring\\ Removal only; Meth. for Method. Unit of $\Delta_{\rm{MAX}}$: HU.}}
	\end{tabular}
	\label{table:simutable}
\end{table}

\subsection{Compatibility between ETB-Cal Algorithm and Pileup Correction}
\begin{figure}[!t]
	\centering
	\centerline{\includegraphics[width=\columnwidth]{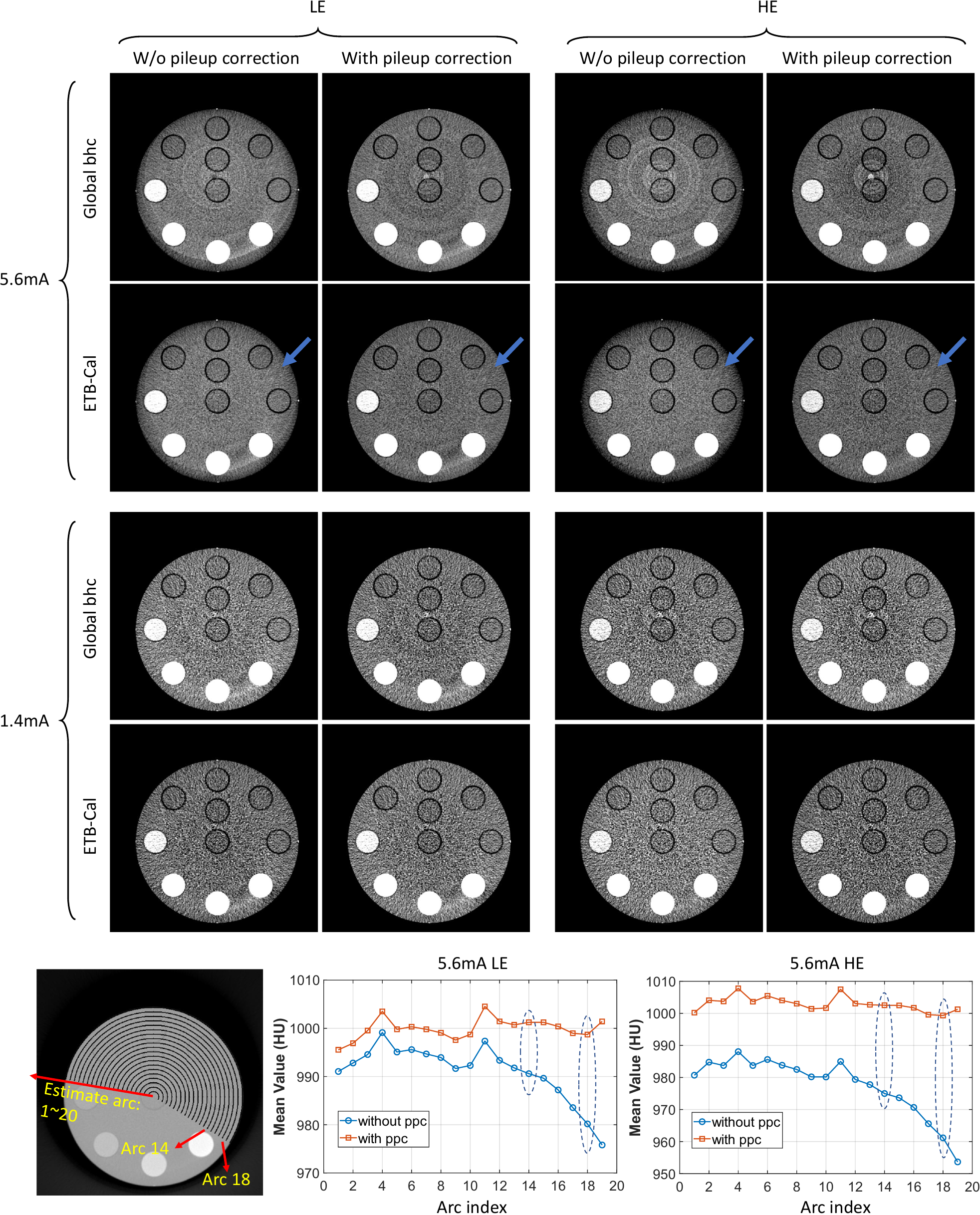}}
	\caption{Effectiveness of ETB-Cal with and without pileup correction, under different tube current (5.6 mA, 1.4 mA), with AC on. To quantify the radial non-uniformity of the phantom caused by pileup, we divided the region from the center to the edge into 20 circular arcs. The average values of all data points on each of the 20 arcs under 5.6 mA tube current were plotted at the bottom. LE: [25, 120] keV, HE: [45, 120] keV, display window width: 120 HU. ppc: pileup correction.}
	\label{fig:Correction with Pileup}
\end{figure}

Fig. \ref{fig:Correction with Pileup} showed the effectiveness of ETB-Cal, with and without pileup correction, at tube currents of 5.6 mA and 1.4 mA, with AC ON. ETB-Cal consistently mitigated ring artifacts irrespective of the pileup-correction setting, as evidenced in the reconstructed images by more uniform gray levels within the water region of the MEPT and by the absence of visible ring artifacts.

Twenty concentric arcs, spanning from the center to the periphery, were selected in the reconstructed images, and the mean pixel value along each arc was analyzed. The results at a tube current of 5.6 mA are shown in the bottom panel of Fig. \ref{fig:Correction with Pileup} for visual comparison. In addition, quantitative analysis of the relative differences in mean values between arcs 18 and 14 across the four tube‑current settings was reported in Table \ref{table:pptable}, indicating that the combined approach maintained its ring‑artifact reduction capability while simultaneously mitigating pileup‑induced decreases in CT numbers at object edges.

While the pileup effects observed in this study were limited due to the experimental limitation of maximum tube current, which precluded evaluation under severe pileup conditions, the results nevertheless offered valid evidence for the compatibility between ETB-Cal and pileup correction methods. Importantly, this sequential correction preserved the original performance of ETB-Cal without compromise, confirming their operational compatibility in scenarios involving both spectral inconsistency and moderate pileup effects. These findings suggested implementation possibilities for practical applications, although additional studies will be required to assess performance under more pronounced pileup conditions.

\section{Discussions \& Conclusions}

In summary, we proposed a novel physics-model based method to correct for spectral inconsistency by modeling it through two terms: (1) a fixed spectral skew term (energy threshold-independent filtration function) determined at a given energy threshold, and (2) a variable energy-threshold bias term that can be directly calculated by using our spectral model as the threshold changes.
After the two terms being computed out in the calibration stage, they will be incorporated into our spectral model to adaptively generate the spectral correction vectors as well as the material decomposition vectors if needed, pixel-by-pixel for PCCT projection data.
Using a minimum set of parameters with explicit physics meaning, such an energy-threshold bias calculator (ETB-Cal) has advantages of computational efficiency, robustness in implementation, and convenience with no need of X-ray fluorescence materials in calibration.
To validate our method, both numerical simulations and physical experiments were carried out on a tabletop PCCT system, with preliminary results showing that the non-uniformity caused by spectral inconsistency can be decreased from 29.3 to 5.8 HU for Gammex MEPT, and from 27.9 to 3.2 HU for Kyoto head phantom, respectively. In contrast, in the same phantom studies, non-uniformity was reduced to 8.3 HU and 9.5 HU by a polynomial-involving model-based approach with no explicit modeling and calculating of energy threshold bias, respectively. In addition, the accuracy of iodine concentration in material decomposition improved significantly after applying our spectral model and ETB-Cal for spectral inconsistency correction.

\begin{table}[t!]
	\centering
	\caption{The average value difference between the 18th arc and that on the 14th arc in Fig. \ref{fig:Correction with Pileup} under different tube current.}
	\begin{tabular}{ccccc}
		\specialrule{1pt}{0pt}{0pt}
		\diagbox{}{} & \multicolumn{2}{c}{\textbf{LE}} & \multicolumn{2}{c}{\textbf{HE}} \\ \hline
		\textbf{Current} & w/o ppc & with ppc & w/o ppc & with ppc \\ \hline 
		5.6 mA & 1.05 \%   & 0.25 \%   & 1.42 \%  & 0.32 \%  \\ 
		4.0 mA & 0.81 \%   & 0.22 \%   & 1.14 \%  & 0.33 \%  \\ 
		2.8 mA & 0.54 \%  & 0.15 \%    & 0.77 \%  & 0.21 \%    \\ 
		1.4 mA & 0.15 \%  & -0.04 \% & 0.38 \% & 0.10 \%   \\ 
		\specialrule{1pt}{0pt}{0pt}
	\end{tabular}
	\label{table:pptable}
\end{table}

As shown in Fig. \ref{fig:Simu validation multi-mat}, among the selected inconsistency-introducing materials, using CdTe and Al as the basis materials for the spectral skew term yielded favorable results in most cases. However, when PMMA was employed as the inconsistency-introducing material, the performance degraded significantly, possibly due to the substantial difference in X-ray attenuation characteristics between PMMA and the two basis materials. This issue could be effectively resolved by incorporating PMMA into the basis materials for spectral skew calculation. Moreover, $\Delta T_i$ in subfig. (b) and $\Delta D_i$ in subfig. (e) did not fully converge to zero. This is caused by approximate calculations performed on discretely defined E-bin intervals.

\begin{figure}
	\centering
	\centerline{\includegraphics[width=1\columnwidth]{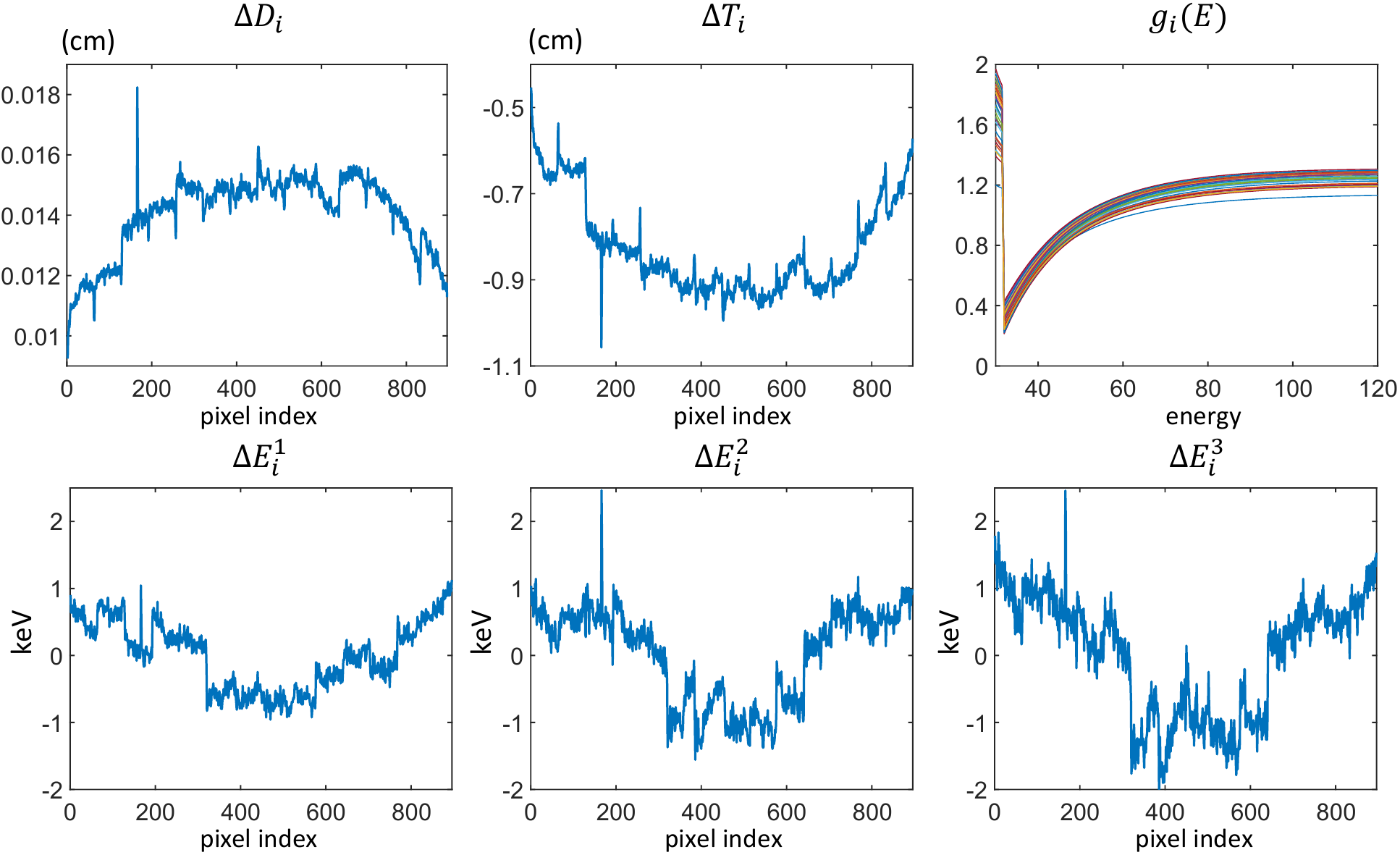}}
	\caption{Values of $\Delta D_i$, $\Delta T_i$, $\Delta E_i^k$ and spectral skew term $g_i(E)$ obtained in the experiment of MEPT: Scan Set-II (reconstructed images shown in Fig. \ref{fig:ETB-cal at different TH} and scan protocol shown in Table \ref{table:parameter}).}
	\label{fig:giE plot}
\end{figure}

In the multi-energy phantom study, we analyzed the results of spectral model parameters obtained in the experiment of Scan Set-II (results shown in Fig. \ref{fig:ETB-cal at different TH} and scan protocol shown in Table \ref{table:parameter}). The results shown in Fig. \ref{fig:giE plot} revealed several key findings: First, the distribution widths of $\Delta D_i$ ($\thicksim$ 0.05 mm) and $\Delta T_i$ ($\thicksim$ 0.4 cm) at $E_T^0$ demonstrated distinct characteristics, whose local variations reflected spectral inconsistencies, while their global x-offset indicated the overall spectral distortion that directly impacted the accuracy of reconstructed images. Second, the anti-correlated distributions between $\Delta D_i$ and $\Delta T_i$ confirmed their coupled physical relationship. Third, all $\Delta E_i^k$ values at $E_T^{1\thicksim3}$ fell within ± 2.5 keV, matching the pre-correction sigma value (2.44 keV) reported in \cite{Liu_2025}, while exhibiting the expected increasing trend with higher $E_T$, consistent with our theoretical analysis in section \ref{sec:Feasibility Analysis of Decoupling}. These results collectively validated that our calibrated parameters operate within physically realistic ranges.

In the head phantom study, the ETB-Cal method achieved measurable inconsistency correction across all scan sets, though it performed more robustly in AC OFF mode. This observation was consistent with the reports of reduced image uniformity during AC ON acquisitions compared to AC OFF conditions, which was documented for the X-THOR FX30 0.75-mm CdTe detector in \cite{marupudi2025evaluation}. The superior performance in AC OFF mode stemmed from two factors: (1) the inherent uniformity advantage of AC OFF data, and (2) challenges in calibrating the step wedge accurately under strong pileup effects. The latter issue was particularly pronounced when comparing transmission data between highly attenuating head phantoms and thin filters, where pileup severity varies significantly.

Compared with the other two model-based methods and the XRF-based method evaluated in this study, our approach was more convenient to implement. Time savings came from faster data collection and, more importantly, from a simpler setup that reduced repeated filter changes and adjustments. In simulations, ETB-Cal with 5 Al filters achieved performance comparable to SFR with 25 step wedge calibration measurements. In experiments, ETB-Cal with 6 Al filters achieved markedly more significant non-uniformity decreasing in MEPT (Scan set-I) and higher MD accuracy in MEPT (Scan set-MD) than SFR with 14 step wedge measurements. 
The BCT method required 37 measurements yet still yielded insufficient interpolation density for practical applications. 
The XRF-based method faces inherent limitations due to low fluorescence photon yields. Obtaining statistically significant counts across all E-bins requires prolonged acquisition times, compounded by the necessity for threshold scanning with multiple measurements near each fluorescence energy. In our typical ETB-Cal implementation, the Levenberg-Marquardt algorithm \cite{nocedal2006numerical} (MATLAB fsolve function) was used to sequentially compute $g_i(E)$ and $\Delta E_i^k$ under two energy thresholds, with total computation times consistently below 30 seconds.

Future work also includes applying the method to a wider range of PCDs and experimental conditions to further demonstrate its generalizability, exploring the possibility of incorporating pileup effects into the model, and further refining its solution process through deep learning network.

\appendices
\section{The Direct Inversion Method}\label{Direct Inversion Method}

The direct inversion method also utilizes two linearization approximation assumptions outlined in \ref{sec:Key Linearization Assumptions}, but treats the parameters $\alpha_i$, $\Delta D_i$, $\Delta T_i$ in $g_i(E)$ and $\Delta E_i^k$ as unknowns to be solved simultaneously.

In this work, a least-square method will be use to approximately solve this objective function.
\begin{equation}
	\begin{aligned}
		&\Delta E^{k}_{i},\alpha_i,\Delta D_{i},\Delta T_{i}=\arg\min\limits_{}\biggl|\biggl|\frac {A_{1i}^k - \Delta D_i A_{2i}^k - \Delta T_i A_{3i}^k}{\sum_{i=1}^{N} \alpha_i A_{1i}^k}\\
		&-\frac {\Delta E^{k}_{i}(B_{1i}^k - \Delta D_i B_{2i}^k - \Delta T_i B_{3i}^k)}{\sum_{i=1}^{N} \alpha_i A_{1i}^k}-\frac{\tilde{\lambda}^{k,measurement}_{i}}{\alpha_i}\biggl|\biggl|\\
	\end{aligned}
	\label{eq:Direct inverse}
\end{equation}

\noindent assuming relatively small $\Delta E^{k}_{i}$ values. 

To ensure stable convergence for the ill-posed problem, we constrained \( \Delta E^{k}_{i} \) and set a maximum iteration limit. The energy threshold was updated by accumulating \( \Delta E^{k}_{i} \) across iterations. The final solution consisted of the energy threshold \( E^{k}_{T}+\sum_{\mathrm{num = 1}}^{\mathrm{iter_{MAX}}}\Delta E^{k,\mathrm{num}}_{i} \), along with the \( \Delta D_i \), \( \Delta T_i \), and \( \alpha_i \) calculated in the last iteration.

Once these parameters are computed out, they will be used in the spectral correction as described in Section \ref{sec:Spectral Inconsistency Correction}.

\section{The sensitivity of $g_i(E)$ to the deviations in $E_T^k$}\label{giE compensation}
If there is a small deviation ($\Delta$) in \( E_T^k \), it will be propagated into \( g_i(E) \). For the integral part associated with ${E^\prime}$ in Eq. \eqref{eq:TTF-spectralModel}:
\begin{equation}\label{eq:inte split} 
	\begin{split}
		\int_{E_{T}^k+\Delta}^\infty{{R_i(E^\prime;E)}{\rm d}E^\prime}
		=C(E)(\int_{E_{T}^k}^\infty{{R_i(E^\prime;E)}{\rm d}E^\prime})
	\end{split}
\end{equation}
where 
\begin{equation}\label{eq:CE} 
	\begin{split}
		C(E)=1-\frac{\int_{E_{T}^k}^{E_{T}^k+\Delta}{{R_i(E^\prime;E)}{\rm d}E^\prime}}{\int_{E_{T}^k}^\infty{{R_i(E^\prime;E)}{\rm d}E^\prime}}
	\end{split}
\end{equation}

Intuitively, the lower $E_T^k$ is, the less sensitive $C(E)$ is to \textit{$\Delta$}, resulting in a smaller error propagation to $g_i(E)$, except for some positions where \( R_i(E_T^{k1};E) \) is significantly larger than \( R_i(E_T^{k2};E) \) (\( k1 < k2 \)).

Quantitatively, by assuming that $R_i(E^\prime;E)$ can be approximated by a series of Gaussian functions, we derived that
\begin{equation}\label{eq:CE-simp} 
	\begin{split}
		C(E)\sim 1 - E_T^{k} \cdot \Delta
	\end{split}
\end{equation}
taking the conclusions of 
\begin{equation}\label{eq:local mass}
	\int_{x_0}^{x_0 + {\rm d}x} \phi(x) {\rm d}x \approx \phi(x_0) {\rm d}x = \frac{1}{\sqrt{2\pi}} e^{-x_0^2/2} {\rm d}x
\end{equation}
\begin{equation}\label{eq:tail mass}
	\int_{x_0}^{\infty} \phi(x) {\rm d}x = \frac{1}{\sqrt{2\pi}} \frac{e^{-x_0^2/2}}{x_0} \left( 1 - \frac{1}{x_0^2} + \mathcal{O}(x_0^{-4}) \right)
\end{equation}
in which $\phi(x) = \frac{1}{\sqrt{2\pi}} e^{-x^2/2}$ is the standard Gaussian probability density function, and $x_0$ ($x_0 \gg 1$).

It is concluded the use of a lower energy threshold $E_T^0$ reduces error propagation to $g_i(E)$ from neglected $\Delta$.

\section{Implementation details of the two comparative model-based methods}\label{Key implementation details of the two comparative model-based methods}
\subsection{SFR Model Method}  

The SFR method establishes a dual-component model to characterize the PCD response. The spectral sensitivity component is formulated as

\begin{equation}  
	\begin{split}
		S_w(E, \beta) = D_w(E)R(E)\cdot\exp\left(-\sum_{j=1}^{N_\beta} \beta_{wj}(E/E_{\text{max}})^j\right),
	\end{split}
\end{equation}  

\noindent where 
$D_w(E)$ denotes the detector response model, $R(E)$ is the x-ray source spectrum, and $\boldsymbol{\beta}$ coefficients provide spectral shape adjustment through a polynomial correction term.

The nonlinear fluence component models the flux-dependent distortions through a polynomial relationship between transmitted ($T^{\text{trans}}$) and measured ($T^{\text{meas}}$) fractional fluence,

\begin{equation}
	T_{w\ell}^{\text{meas}} = T_{w\ell}^{\text{trans}} + \sum_{k=1}^{N_\alpha} \alpha_{wk}(T_{w\ell}^{\text{trans}})^k,
\label{eq:Talpha}
\end{equation}

The complete model parameters are determined through an optimization framework that minimizes the Kullback-Leibler divergence between measured and modeled transmission data, incorporating Tikhonov regularization for stability,

\begin{equation}  
	\begin{split}
		&\alpha^*, \beta^* = \\
		&\underset{\alpha,\beta}{\mathrm{arg\,min}} \left\{ \sum_\ell D_{\text{KL}}\left(T_w^{\text{data}} ,\text{max}(T_w^{\text{meas}}(\alpha,\beta),\epsilon)\right) + \frac{\gamma}{2} \|\beta\|_2^2 \right\},
	\end{split}
\end{equation}  

\noindent in which $\gamma$ is the regularization parameter to maintain model fidelity while preventing overfitting.

\subsection{BCT Model Method}

The method employs a mathematical formulation that begins with the approximation of energy-dependent x-ray transmittance. The x-ray transmittance $X_j(E)$ for the $j$-th line integral is represented as
\begin{equation}
	X_j(E) \approx \sum_{i=0}^{k-1} \theta_{(i,j)} P_i(E),
\end{equation}
\noindent where $P_i(E)$ are orthogonal polynomials (typically Gram polynomials), and $\theta_{(i,j)}$ are the coefficients to be estimated. The coefficients are determined through a penalized maximum likelihood estimation
\begin{equation}
	\hat{\mathbf{\theta}}_j = \arg\min_{\theta} \| y_j - B\theta \|_2^2 + \lambda \| K\theta \|_2^2.
\end{equation}
\noindent Here, $B$ represents the sensing matrix derived from the SRE model and incident spectrum, $\lambda$ is a regularization parameter, and $K$ is a contrast matrix. The solution takes the closed-form expression,
\begin{equation}
	\hat{\mathbf{\theta}}_j = (B^T B + \lambda K^T K)^{-1} B^T y_j.
\end{equation}
The basis line-integrals $\mathbf{v}_j$ are subsequently estimated through least squares fitting as
\begin{equation}
	\hat{\mathbf{v}}_j = \arg\min_{\mathbf{v}} \| \ln(\hat{X}_j(E)) + \Phi(E)\mathbf{v} \|_d^2,
\end{equation}

\noindent where $\Phi(E)$ represents the basis functions for photoelectric and Compton scattering effects. The final step involves an iterative bias correction procedure to compensate for approximation errors.

For each combination $\mathbf{v} \in \Omega$, where $\Omega$ represents the predefined domain of practical attenuation values, multiple calibration data are generated through either physical measurements or Monte Carlo simulations.

Beginning with the primitive table $BCT^{(1)}(\mathbf{v})$, subsequent tables are generated through recursive application of 
\begin{equation}
	BCT^{(q+1)}(\mathbf{v}) = \mathbb{E}[\hat{\mathbf{v}}^{(q)}] - \mathbf{v}
\end{equation}

\noindent where $\hat{\mathbf{v}}^{(q)}$ represents estimates obtained using the $q$-th iteration's correction table. The iterative process continues until the residual bias falls below a predetermined threshold or reaches a maximum iteration count.

\section{Pileup Correction Compatibility}\label{Pileup Correction Compatibility}
While our ETB-Cal model does not inherently include pileup effect and its correction, we demonstrate its compatibility with conventional pileup mitigation approaches, which can be written as
\begin{equation}
	\begin{split}
	p_{c}=&f_{ETB-Cal}(\mathcal{P}(f_{ppc}(Cnt_{phantom})),\\
	\end{split}
	\label{eq:Compatibility} 
\end{equation}
where $Cnt$ denotes the raw measured counts without any corrections, $f_{ppc}$ represents the pileup correction applied to the count values, $\mathcal{P}$ refers to negative logarithm of the transmission data, $f_{ETB-Cal}$ signifies ETB-Cal correction, and $p_c$ represents the final corrected projections that comprehensively address pileup effects, spectral inconsistencies, and beam hardening artifacts through the complete correction pipeline.

Our implementation followed a two-stage process: 

STAGE I: We implemented a hybrid pileup correction approach utilizing current-count curve analysis\cite{lee2000new} to approximate the pileup correction function $f_{ppc}$, thereby effectively modeling the nonlinear detector response under high flux conditions. The hybrid model is written as
\begin{equation}
	m=\frac{n\cdot\rm{e}^{-n\tau_p}}{1+n\tau_n},
	\label{eq:hybrid pileup correction} 
\end{equation}
makes use of the paralyzable deadtime $\tau_p$ and the nonparalyzable deadtime $\tau_n$, in which $m$ and $n$ respectively represent the observed count rate and the true count rate.

STAGE II: After applying the pileup compensation to the raw detector counts, we implemented the established ETB-Calibration workflow (section \ref{sec:ETB-Cal Decomposition Algorithm} and \ref{sec:Spectral Inconsistency Correction}), which performs spectral correction through energy-threshold-based calibration to mitigate spectral inconsistency. 

This pipeline ultimately yielded both the ETB-Cal inconsistency parameters and corrected spectral CT reconstruction images of the phantom.

\section*{Acknowledgments}
The authors are grateful to Dr. Adam Wang from Stanford University for in-depth discussions on spectral modeling and corrections, and to Dr. Binxiang Qi and Yukang Song from Tsinghua University for their generous helps in experiments. Sincere appreciations also go to the editors and anonymous reviewers for their constructive comments and insightful suggestions.

\bibliographystyle{MedPhys}
\bibliography{ref}

@article{alvarez2011estimator,
	title={Estimator for photon counting energy selective x-ray imaging with multibin pulse height analysis},
	author={Alvarez, Robert E},
	journal={Medical physics},
	volume={38},
	number={5},
	pages={2324--2334},
	year={2011},
	publisher={Wiley Online Library}
}

@article{Liu_2025,
	doi = {10.1088/1748-0221/20/06/P06044},
	url = {https://dx.doi.org/10.1088/1748-0221/20/06/P06044},
	year = {2025},
	month = {jun},
	publisher = {IOP Publishing},
	volume = {20},
	number = {06},
	pages = {P06044},
	author = {Liu, Yi and others},
	title = {Threshold equalization and energy calibration of a novel multi-threshold photon counting detector},
	journal = {Journal of Instrumentation},
}

@article{Vespucci2019RobustEcal,
	title={Robust energy calibration technique for photon counting spectral detectors},
	author={Vespucci, Stefano and Park, Chan Soo and Torrico, Raul and Das, Mini},
	journal={IEEE transactions on medical imaging},
	volume={38},
	number={4},
	pages={968--978},
	year={2018},
	publisher={IEEE}
}

@article{li2016feasible,
	title={Feasible energy calibration for multi-threshold photon-counting detectors based on reconstructed XRF spectra},
	author={Li, Ruizhe and Li, Liang and Chen, Zhiqiang},
	journal={IEEE Transactions on Radiation and Plasma Medical Sciences},
	volume={1},
	number={2},
	pages={109--120},
	year={2016},
	publisher={IEEE}
}

@article{lee2000new,
	title={A new G--M counter dead time model},
	author={Lee, Sang Hoon and Gardner, Robin P},
	journal={Applied Radiation and Isotopes},
	volume={53},
	number={4-5},
	pages={731--737},
	year={2000},
	publisher={Elsevier}
}

@article{abramowitz1972handbook,
	title={Handbook of Mathematical Functions with Formulas, Graphs, and Mathematical Tables. National Bureau of Standards Applied Mathematics Series 55. Tenth Printing.},
	author={Abramowitz, Milton and Stegun, Irene A},
	year={1972},
	publisher={ERIC}
}

@article{broennimann2006pilatus,
	title={The PILATUS 1M detector},
	author={Broennimann, Ch and others},
	journal={Synchrotron Radiation},
	volume={13},
	number={2},
	pages={120--130},
	year={2006},
	publisher={International Union of Crystallography}
}

@article{biguri2016tigre,
	title={TIGRE: a MATLAB-GPU toolbox for CBCT image reconstruction},
	author={Biguri, Ander and Dosanjh, Manjit and Hancock, Steven and Soleimani, Manuchehr},
	journal={Biomedical Physics \& Engineering Express},
	volume={2},
	number={5},
	pages={055010},
	year={2016},
	publisher={IOP Publishing}
}

@article{marupudi2025evaluation,
	title={Evaluation of charge summing correction in CdTe-based photon-counting detectors for breast CT: performance metrics and image quality},
	author={Marupudi, Sriharsha and Manus, Joseph A and Ghani, Muhammad U and Glick, Stephen J and Ghammraoui, Bahaa},
	journal={Journal of Medical Imaging},
	volume={12},
	number={1},
	pages={013501--013501},
	year={2025},
	publisher={Society of Photo-Optical Instrumentation Engineers}
}

@inproceedings{ullberg2018photon,
  title={Photon counting dual energy x-ray imaging at CT count rates: measurements and implications of in-pixel charge sharing correction},
  author={Ullberg, Christer and Urech, Mattias and Eriksson, Charlotte and Stewart, Alexander and Weber, Niclas},
  booktitle={Medical imaging 2018: physics of medical imaging},
  volume={10573},
  pages={316--323},
  year={2018},
  organization={SPIE}
}

@article{wang2023dual,
  title={Dual-energy head cone-beam CT using a dual-layer flat-panel detector: hybrid material decomposition and a feasibility study},
  author={Wang, Zhilei and others},
  journal={Medical Physics},
  volume={50},
  number={11},
  pages={6762--6778},
  year={2023},
  publisher={Wiley Online Library}
}

@article{sijbers2004reduction,
  title={Reduction of ring artefacts in high resolution micro-CT reconstructions},
  author={Sijbers, Jan and Postnov, Andrei},
  journal={Physics in Medicine \& Biology},
  volume={49},
  number={14},
  pages={N247},
  year={2004},
  publisher={IOP Publishing}
}

@article{sidky2005robust,
  title={A robust method of x-ray source spectrum estimation from transmission measurements: Demonstrated on computer simulated, scatter-free transmission data},
  author={Sidky, Emil Y and Yu, Lifeng and Pan, Xiaochuan and Zou, Yu and Vannier, Michael},
  journal={Journal of applied physics},
  volume={97},
  number={12},
  year={2005},
  publisher={AIP Publishing}
}

@book{hsieh2022computed,
	author    = {Hsieh Jiang},
	title     = {Computed Tomography: Principles, Design, Artifacts, and Recent Advances},
	edition   = {4},
	publisher = {SPIE Press},
	year      = {2022},
	address   = {Bellingham, WA}
}

@book{nocedal2006numerical,
	title={Numerical optimization},
	author={Nocedal, Jorge and Wright, Stephen J},
	year={2006},
	publisher={Springer}
}

@article{runge1901empirische,
	title={{\"U}ber empirische Funktionen und die Interpolation zwischen {\"a}quidistanten Ordinaten},
	author={Runge, Carl and others},
	journal={Zeitschrift f{\"u}r Mathematik und Physik},
	volume={46},
	number={224-243},
	pages={20},
	year={1901}
}

@misc{FX35,
	title        = {DC-Hydra - Varex Imaging},
	howpublished = {https://www.vareximaging.com/solutions/dc-hydra/},
	note         = {Accessed: 2025-11-01}
}

@inproceedings{chen2024physics,
  title={Physics-based modeling of energy threshold induced spectral inconsistency and its adaptive correction scheme for photon-counting CT},
  author={Chen, Yuting and Xing, Yuxiang and Gao, Hewei},
  booktitle={Medical Imaging 2024: Physics of Medical Imaging},
  volume={12925},
  pages={6--10},
  year={2024},
  organization={SPIE}
}

@article{rodesch2023comparison,
	title={Comparison of threshold energy calibrations of a photon-counting detector and impact on CT reconstruction},
	author={Rodesch, Pierre-Antoine and Richtsmeier, Devon and Guliyev, Elmaddin and Iniewski, Kris and Bazalova-Carter, Magdalena},
	journal={IEEE Transactions on Radiation and Plasma Medical Sciences},
	volume={7},
	number={3},
	pages={263--272},
	year={2023},
	publisher={IEEE}
}

@article{ballabriga2021review,
	author = {Ballabriga, R. and others},
	title = {Photon Counting Detectors for X-Ray Imaging With Emphasis on CT},
	journal = {IEEE Transactions on Radiation and Plasma Medical Sciences},
	volume = {5},
	number = {4},
	pages = {422-440},
	ISSN = {2469-7311
	2469-7303},
	DOI = {10.1109/trpms.2020.3002949},
	year = {2021},
	type = {Journal Article}
}

@article{danielsson2021review,
	author = {Danielsson, M. and Persson, M. and Sjolin, M.},
	title = {Photon-counting x-ray detectors for CT},
	journal = {Phys Med Biol},
	volume = {66},
	number = {3},
	pages = {03TR01},
	ISSN = {1361-6560 (Electronic)
	0031-9155 (Linking)},
	DOI = {10.1088/1361-6560/abc5a5},
	url = {https://www.ncbi.nlm.nih.gov/pubmed/33113525},
	year = {2021},
	type = {Journal Article}
}

@article{Flohr2020review,
	author = {Flohr, T. and Petersilka, M. and Henning, A. and Ulzheimer, S. and Ferda, J. and Schmidt, B.},
	title = {Photon-counting CT review},
	journal = {Phys Med},
	volume = {79},
	pages = {126-136},
	ISSN = {1724-191X (Electronic)
	1120-1797 (Linking)},
	DOI = {10.1016/j.ejmp.2020.10.030},
	url = {https://www.ncbi.nlm.nih.gov/pubmed/33249223},
	year = {2020},
	type = {Journal Article}
}

@article{Flohr2020photoncounting,
	author = {Flohr, Thomas and Ulzheimer, Stefan and Petersilka, Martin and Schmidt, Bernhard},
	title = {Basic principles and clinical potential of photon-counting detector CT},
	journal = {Chinese Journal of Academic Radiology},
	volume = {3},
	number = {1},
	pages = {19-34},
	ISSN = {2520-8985
	2520-8993},
	DOI = {10.1007/s42058-020-00029-z},
	year = {2020},
	type = {Journal Article}
}

@article{lee2018spekdisto,
	author = {Lee, O. and Kappler, S. and Polster, C. and Taguchi, K.},
	title = {Estimation of Basis Line-Integrals in a Spectral Distortion-Modeled Photon Counting Detector Using Low-Order Polynomial Approximation of X-ray Transmittance},
	journal = {IEEE Trans Med Imaging},
	volume = {36},
	number = {2},
	pages = {560-573},
	ISSN = {1558-254X (Electronic)
	0278-0062 (Linking)},
	DOI = {10.1109/TMI.2016.2621821},
	url = {https://www.ncbi.nlm.nih.gov/pubmed/27810801},
	year = {2017},
	type = {Journal Article}
}

@article{Rajendran2021review,
	author = {Rajendran, K. and others},
	title = {Full field-of-view, high-resolution, photon-counting detector CT: technical assessment and initial patient experience},
	journal = {Phys Med Biol},
	volume = {66},
	number = {20},
	ISSN = {1361-6560 (Electronic)
	0031-9155 (Print)
	0031-9155 (Linking)},
	DOI = {10.1088/1361-6560/ac155e},
	url = {https://www.ncbi.nlm.nih.gov/pubmed/34271558},
	year = {2021},
	type = {Journal Article}
}

@article{Shikhaliev2011review,
	author = {Shikhaliev, P. M. and Fritz, S. G.},
	title = {Photon counting spectral CT versus conventional CT: comparative evaluation for breast imaging application},
	journal = {Phys Med Biol},
	volume = {56},
	number = {7},
	pages = {1905-30},
	ISSN = {1361-6560 (Electronic)
	0031-9155 (Linking)},
	DOI = {10.1088/0031-9155/56/7/001},
	url = {https://www.ncbi.nlm.nih.gov/pubmed/21364268},
	year = {2011},
	type = {Journal Article}
}

@article{RN45,
	author = {Shikhaliev, P. M. and Xu, T. and Molloi, S.},
	title = {Photon counting computed tomography: concept and initial results},
	journal = {Med Phys},
	volume = {32},
	number = {2},
	pages = {427-36},
	ISSN = {0094-2405 (Print)
	0094-2405 (Linking)},
	DOI = {10.1118/1.1854779},
	url = {https://www.ncbi.nlm.nih.gov/pubmed/15789589},
	year = {2005},
	type = {Journal Article}
}

@article{sidky2022spekcali,
	author = {Sidky, E. Y. and Paul, E. R. and Gilat-Schmidt, T. and Pan, X.},
	title = {Spectral calibration of photon-counting detectors at high photon flux},
	journal = {Med Phys},
	volume = {49},
	number = {10},
	pages = {6368-6383},
	ISSN = {2473-4209 (Electronic)
	0094-2405 (Print)
	0094-2405 (Linking)},
	DOI = {10.1002/mp.15942},
	url = {https://www.ncbi.nlm.nih.gov/pubmed/35975670},
	year = {2022},
	type = {Journal Article}
}

@article{Taguchi2017review,
	author = {Taguchi, K.},
	title = {Energy-sensitive photon counting detector-based X-ray computed tomography},
	journal = {Radiol Phys Technol},
	volume = {10},
	number = {1},
	pages = {8-22},
	ISSN = {1865-0341 (Electronic)
	1865-0333 (Linking)},
	DOI = {10.1007/s12194-017-0390-9},
	url = {https://www.ncbi.nlm.nih.gov/pubmed/28138947},
	year = {2017},
	type = {Journal Article}
}

@article{Willemink2018review,
	author = {Willemink, M. J. and Persson, M. and Pourmorteza, A. and Pelc, N. J. and Fleischmann, D.},
	title = {Photon-counting CT: Technical Principles and Clinical Prospects},
	journal = {Radiology},
	volume = {289},
	number = {2},
	pages = {293-312},
	ISSN = {1527-1315 (Electronic)
	0033-8419 (Linking)},
	DOI = {10.1148/radiol.2018172656},
	url = {https://www.ncbi.nlm.nih.gov/pubmed/30179101},
	year = {2018},
	type = {Journal Article}
}

@article{yu2016photoncounting,
	author = {Yu, Z. and others},
	title = {Evaluation of conventional imaging performance in a research whole-body CT system with a photon-counting detector array},
	journal = {Phys Med Biol},
	volume = {61},
	number = {4},
	pages = {1572-95},
	ISSN = {1361-6560 (Electronic)
	0031-9155 (Print)
	0031-9155 (Linking)},
	DOI = {10.1088/0031-9155/61/4/1572},
	url = {https://www.ncbi.nlm.nih.gov/pubmed/26835839},
	year = {2016},
	type = {Journal Article}
}

@article{Mohamed2021kedge,
	author = {Si-Mohamed, S. A. and others},
	title = {In Vivo Molecular K-Edge Imaging of Atherosclerotic Plaque Using Photon-counting CT},
	journal = {Radiology},
	volume = {300},
	number = {1},
	pages = {98-107},
	ISSN = {1527-1315 (Electronic)
	0033-8419 (Print)
	0033-8419 (Linking)},
	DOI = {10.1148/radiol.2021203968},
	url = {https://www.ncbi.nlm.nih.gov/pubmed/33944628},
	year = {2021},
	type = {Journal Article}
}

@article{Roessl2011kedge,
	author = {Roessl, E. and Brendel, B. and Engel, K. J. and Schlomka, J. P. and Thran, A. and Proksa, R.},
	title = {Sensitivity of photon-counting based K-edge imaging in X-ray computed tomography},
	journal = {IEEE Trans Med Imaging},
	volume = {30},
	number = {9},
	pages = {1678-90},
	ISSN = {1558-254X (Electronic)
	0278-0062 (Linking)},
	DOI = {10.1109/TMI.2011.2142188},
	url = {https://www.ncbi.nlm.nih.gov/pubmed/21507770},
	year = {2011},
	type = {Journal Article}
}

@article{Gimenez2011chargesharing,
   author = {Gimenez, E. N. and others},
   title = {Study of charge-sharing in MEDIPIX3 using a micro-focused synchrotron beam},
   journal = {Journal of Instrumentation},
   volume = {6},
   number = {01},
   pages = {C01031-C01031},
   ISSN = {1748-0221},
   DOI = {10.1088/1748-0221/6/01/c01031},
   year = {2011},
   type = {Journal Article}
}

@article{Ding2014xrfR,
	author = {Ding, H. and Cho, H. M. and Barber, W. C. and Iwanczyk, J. S. and Molloi, S.},
	title = {Characterization of energy response for photon-counting detectors using x-ray fluorescence},
	journal = {Med Phys},
	volume = {41},
	number = {12},
	pages = {121902},
	ISSN = {2473-4209 (Electronic)
	0094-2405 (Print)
	0094-2405 (Linking)},
	DOI = {10.1118/1.4900820},
	url = {https://www.ncbi.nlm.nih.gov/pubmed/25471962},
	year = {2014},
	type = {Journal Article}
}

@article{Ge2017kedge,
	author = {Ge, Y. and Ji, X. and Zhang, R. and Li, K. and Chen, G. H.},
	title = {K-edge energy-based calibration method for photon counting detectors},
	journal = {Phys Med Biol},
	volume = {63},
	number = {1},
	pages = {015022},
	ISSN = {1361-6560 (Electronic)
	0031-9155 (Print)
	0031-9155 (Linking)},
	DOI = {10.1088/1361-6560/aa9637},
	url = {https://www.ncbi.nlm.nih.gov/pubmed/29072192},
	year = {2017},
	type = {Journal Article}
}

@article{fangwei2020NN,
	author = {Fang, Wei and Li, Liang and Chen, Zhiqiang},
	title = {Removing Ring Artefacts for Photon-Counting Detectors Using Neural Networks in Different Domains},
	journal = {IEEE Access},
	volume = {8},
	pages = {42447-42457},
	ISSN = {2169-3536},
	DOI = {10.1109/access.2020.2977096},
	year = {2020},
	type = {Journal Article}
}

@article{Panta2015calibration,
   author = {Panta, Raj Kumar and Walsh, Michael F. and Bell, Stephen T. and Anderson, Nigel G. and Butler, Anthony P. and Butler, Philip H.},
   title = {Energy Calibration of the Pixels of Spectral X-ray Detectors},
   journal = {IEEE Transactions on Medical Imaging},
   volume = {34},
   number = {3},
   pages = {697-706},
   ISSN = {0278-0062
1558-254X},
   DOI = {10.1109/tmi.2014.2337881},
   year = {2015},
   type = {Journal Article}
}

@article{lee2018kedge,
	author = {Lee, O. and Kappler, S. and Polster, C. and Taguchi, K.},
	title = {Estimation of Basis Line-Integrals in a Spectral Distortion-Modeled Photon Counting Detector Using Low-Rank Approximation-Based X-Ray Transmittance Modeling: K-Edge Imaging Application},
	journal = {IEEE Trans Med Imaging},
	volume = {36},
	number = {11},
	pages = {2389-2403},
	ISSN = {1558-254X (Electronic)
	0278-0062 (Linking)},
	DOI = {10.1109/TMI.2017.2746269},
	url = {https://www.ncbi.nlm.nih.gov/pubmed/28866486},
	year = {2017},
	type = {Journal Article}
}

@article{Rajbhandary2018calibration,
	author = {Rajbhandary, P. L. and Hsieh, S. S. and Pelc, N. J.},
	title = {Effect of Spectral Degradation and Spatio-Energy Correlation in X-Ray PCD for Imaging},
	journal = {IEEE Trans Med Imaging},
	volume = {37},
	number = {8},
	pages = {1910-1919},
	ISSN = {1558-254X (Electronic)
	0278-0062 (Linking)},
	DOI = {10.1109/TMI.2018.2834369},
	url = {https://www.ncbi.nlm.nih.gov/pubmed/29993882},
	year = {2018},
	type = {Journal Article}
}

@article{Iniewski2007chargesharing,
   author = {Iniewski, K. and Chen, H. and Bindley, G. and Kuvvetli, I. and Budtz-Jorgensen, C.},
   title = {Modeling charge-sharing effects in pixellated CZT detectors},
   journal = {2007 Ieee Nuclear Science Symposium Conference Record, Vols 1-11},
   pages = {4608-+},
   ISSN = {1095-7863},
   DOI = {Doi 10.1109/Nssmic.2007.4437135},
   url = {<Go to ISI>://WOS:000257380403161},
   year = {2007},
   type = {Journal Article}
}

@article{Kafaee2020pileup,
   author = {Kafaee, Mahdi and Goodarzi, Mohammad Mohsen},
   title = {Pile-Up Correction in Spectroscopic Signals Using Regularized Sparse Reconstruction},
   journal = {IEEE Transactions on Nuclear Science},
   volume = {67},
   number = {5},
   pages = {858-862},
   ISSN = {0018-9499
1558-1578},
   DOI = {10.1109/tns.2020.2985104},
   year = {2020},
   type = {Journal Article}
}

@article{Cammin2014distortions,
	author = {Cammin, Jochen and Xu, Jennifer and Barber, William C. and Iwanczyk, Jan S. and Hartsough, Neal E. and Taguchi, Katsuyuki},
	title = {A cascaded model of spectral distortions due to spectral response effects and pulse pileup effects in a photon‐counting x‐ray detector for CT},
	journal = {Medical Physics},
	volume = {41},
	number = {4},
	ISSN = {0094-2405
	2473-4209},
	DOI = {10.1118/1.4866890},
	year = {2014},
	type = {Journal Article}
}

@article{Behbahani2020pileup,
   author = {Mohammadian-Behbahani, Mohammad-Reza and Saramad, Shahyar},
   title = {A comparison study of the pile-up correction algorithms},
   journal = {Nuclear Instruments and Methods in Physics Research Section A: Accelerators, Spectrometers, Detectors and Associated Equipment},
   volume = {951},
   ISSN = {01689002},
   DOI = {10.1016/j.nima.2019.163013},
   year = {2020},
   type = {Journal Article}
}

@article{alvarez1976energy,
  title={Energy-selective reconstructions in x-ray computerised tomography},
  author={Alvarez, Robert E and Macovski, Albert},
  journal={Physics in Medicine \& Biology},
  volume={21},
  number={5},
  pages={733},
  year={1976},
  publisher={IOP Publishing}
}

\end{document}